\documentclass[twocolumn]{aastex62}

\newcommand{\cas}{Cas~A}
\newcommand{\sna}{SN~1987A}

\newcommand{\iib}{IIb}

\newcommand{\henv}{0.5}
\newcommand{\kmps}{~km~s$^{-1}$}

\newcommand{\Msun}{~M$_{\sun}$}
\newcommand{\Lsun}{~L$_{\sun}$}
\newcommand{\hi}{H\textsc{\,i}}

\usepackage{amsmath}

\received{May 4, 2018}
\revised{January 0, 0000}
\accepted{\today}
\submitjournal{ApJ}
\shorttitle{X-ray Absorption in Young CC SNRs}
\shortauthors{Alp et al.}

\begin{document}
\title{X-ray Absorption in Young Core-Collapse Supernova Remnants}

\correspondingauthor{Dennis Alp}
\email{dalp@kth.se}

\author[0000-0002-0427-5592]{Dennis Alp}
\affil{Department of Physics, KTH Royal Institute of Technology, 
  The Oskar Klein Centre, AlbaNova, SE\nobreakdash{-}106 91 Stockholm, Sweden}

\author[0000-0003-0065-2933]{Josefin Larsson}
\affil{Department of Physics, KTH Royal Institute of Technology, 
  The Oskar Klein Centre, AlbaNova, SE\nobreakdash{-}106 91 Stockholm, Sweden}

\author[0000-0001-8532-3594]{Claes Fransson}
\affil{Department of Astronomy, Stockholm University, 
  The Oskar Klein Centre, AlbaNova, SE\nobreakdash{-}106 91 Stockholm, Sweden}

\author[0000-0002-1663-4513]{Michael Gabler}
\affil{Max Planck Institute for Astrophysics, 
       Karl-Schwarzschild-Str.~1, D\nobreakdash{-}85748 Garching, Germany}

\author[0000-0001-8400-8891]{Annop Wongwathanarat}
\affil{Max Planck Institute for Astrophysics, 
       Karl-Schwarzschild-Str.~1, D\nobreakdash{-}85748 Garching, Germany}

\author[0000-0002-0831-3330]{Hans-Thomas Janka}
\affil{Max Planck Institute for Astrophysics, 
       Karl-Schwarzschild-Str.~1, D\nobreakdash{-}85748 Garching, Germany}

\begin{abstract}
  The material expelled by core-collapse supernova (SN) explosions
  absorbs X-rays from the central regions. We use SN models based on
  three-dimensional neutrino-driven explosions to estimate optical
  depths to the center of the explosion, compare different progenitor
  models, and investigate the effects of explosion asymmetries. The
  optical depths below 2~keV for progenitors with a remaining hydrogen
  envelope are expected to be high during the first century after the
  explosion due to photoabsorption. A typical optical depth is
  $100\,t_4^{-2}\,E^{-2}$, where $t_4$ is the time since the explosion
  in units of 10\,000 days (${\sim}$27 years) and $E$ the energy in
  units of keV. Compton scattering dominates above 50~keV, but the
  scattering depth is lower and reaches unity already at ${\sim}$1000
  days at 1~MeV. The optical depths are approximately an order of
  magnitude lower for hydrogen-stripped progenitors. The metallicity
  of the SN ejecta is much higher than in the interstellar medium,
  which enhances photoabsorption and makes absorption edges
  stronger. These results are applicable to young SN remnants in
  general, but we explore the effects on observations of \sna{} and
  the compact object in \cas{} in detail. For \sna{}, the absorption
  is high and the X-ray upper limits of ${\sim}$100\Lsun{} on a
  compact object are approximately an order of magnitude less
  constraining than previous estimates using other absorption
  models. The details are presented in an accompanying
  paper~\citep{alp18}. For the central compact object in \cas{}, we
  find no significant effects of our more detailed absorption model on
  the inferred surface temperature.
\end{abstract}


\section{Introduction}\label{sec:intro}
Core-collapse supernova (SN, core-collapse unless otherwise stated)
explosions expel large amounts of material into their
surroundings. The ejecta dominate the absorption for X-rays up to the
young SN remnant (SNR) stage. To accurately analyze X-ray
observations, it is important to estimate the optical depth as a
function of energy along the line of sight. Most X-ray absorption
models~\citep[e.g.][]{wilms00, gatuzz15} have been developed in order
to account for photoabsorption by the interstellar medium (ISM) and
the column densities are often obtained by retrieving the hydrogen
column number density $N_\mathrm{ISM}(\mathrm{H})$ from \hi{}
surveys~\citep{dickey90, kalberla05, winkel16, hi4pi16}. We use the
notation $N_\mathrm{ISM}(\mathrm{H})$ instead of the standard
$N_\mathrm{H}$ to differentiate $N_\mathrm{ISM}(\mathrm{H})$ from
$N_\mathrm{SN}(\mathrm{H})$, which is the hydrogen column number
density of the ejecta. Other methods include letting
$N_\mathrm{ISM}(\mathrm{H})$ be a free parameter that is fitted or
inferring $N_\mathrm{ISM}(\mathrm{H})$ from observations of dust or
molecular lines. It is very common to assume abundances relative to
hydrogen of all other elements to be solar or representative ISM
abundances~\citep[e.g.][]{anders89, grevesse98, wilms00}. The main
difference in young SNRs is that the metallicity is much higher than
in the ISM.

Here, we use three-dimensional (3D) neutrino-driven SN explosion
models to estimate the column densities of all elements that have a
significant contribution to the optical depth. Similar estimates have
been made by other studies using individual progenitors and less
realistic SN models~\citep{fransson87, serafimovich04, orlando15,
  esposito18}. Our results are generally in agreement with previous
works, but we explore 3D SN explosion models, quantify the
uncertainties introduced by SN asymmetries, and compare different
progenitors. The need for more general and detailed ejecta absorption
models have been indicated in the literature~\citep[e.g.][]{stage04,
  shtykovskiy05, long12, posselt13, orlando15}. We focus on the
absorption toward the center of the SN, where the compact object is
expected to reside, and also re-analyze X-ray observations of the
compact objects in \sna{} and \cas{} using our absorption model.

\sna{} is the closest observed SN in more than three
centuries~\citep[for reviews of SN 1987A, see][]{arnett89, mccray93,
  mccray16}. It is located in the Large Magellanic Cloud at a distance
of approximately 50~kpc~\citep{panagia91, panagia99}. \sna{} was a
Type~II\nobreakdash{-}pec core-collapse SN and is therefore expected
to leave a compact object. The observed neutrino burst confirms the
existence of a compact object~\citep{hirata87, hirata88, bionta87,
  bratton88}, but electromagnetic radiation from the compact object
has never been observed despite more than 30 years of searches
spanning the entire observable spectrum. The X-ray absorption below a
few keV in \sna{} is very high at current epochs, which limits our
ability to probe the central parts of the ejecta at these energies. An
accompanying paper~\citep{alp18} uses estimates from the current work
to model the X-ray absorption and presents a detailed study of
multiwavelength flux limits on the compact object. The main result of
the X-ray analysis is that the limits are about 100\Lsun{}, depending
on the assumed spectrum of the compact object. This is approximately
an order of magnitude less constraining than previously published
X-ray limits. The difference compared to previous absorption models is
a substantial increase of the column densities, particularly for the
metals, which dominate the opacity in the relevant 0.3--10~keV energy
range.

\cas{} is a SNR at a distance of $3.4^{+0.3}_{-0.1}$~kpc, based on
proper-motion and age data~\citep{reed95}. \cas{} was created by a
Type~IIb core-collapse SN~\citep{krause08, rest11}. Extrapolation of
the proper motion of the expanding ejecta estimates the explosion date
of the SN to be around the year 1670~\citep{fesen06b}. The SN was
possibly observed by John Flamsteed on 16 August
1680~\citep{flamsteed25, kamper80, hughes80, ashworth80}. The central
compact object (CCO) created by the SN was detected in the first light
images of \textit{Chandra}~\citep{tananbaum99, pavlov00,
  chakrabarty01}. The spectrum of the CCO has been interpreted as
thermal emission from a neutron star (NS) surface with a carbon
atmosphere~\citep{ho09}. Fast cooling of the CCO has been reported
based on analyses of \textit{Chandra} X-ray
observations~\citep{heinke10, elshamouty13}, but the significance of
the cooling has been discussed by \citet{posselt13}.  Here, we simply
repeat part of their analyses and explore the difference introduced by
using a more realistic absorption model based on our 3D SN explosion
models.

This paper is organized as follows. We introduce the 3D SN explosion
models in Section~\ref{sec:mod} and describe the methods in
Section~\ref{sec:methods}. The estimated optical depths and abundances
are presented in Section~\ref{sec:results} and implications and
uncertainties of the results are discussed in
Section~\ref{sec:discussion}. We provide a summary and the main
conclusions in Section~\ref{sec:conclusions}. The investigation of the
effects of ejecta absorption on the interpretation of the observed
\cas{} CCO spectrum is contained in Appendix~\ref{app:cas}.

\section{Models}\label{sec:mod}
\begin{deluxetable*}{lccccccccc}
\caption{Supernova Explosion Models\label{tab:mod}}
  \tablewidth{0pt}
  \tablehead{\colhead{Model\tablenotemark{a}} & \colhead{Name} & \colhead{Type} & \colhead{$M_\mathrm{NS}$} & \colhead{$M_\mathrm{ej}$} & \colhead{$M_\mathrm{tot}$\tablenotemark{b}} & \colhead{$t_\text{late}$} \\
    \colhead{} & \colhead{} & \colhead{} & \colhead{(M$_{\sun}$)} & \colhead{(M$_{\sun}$)} & \colhead{(M$_{\sun}$)} & \colhead{(days)}}\startdata
  B15-1-pw     & B15 & BSG         & 1.2 &            14.2 &            15.4 & 156 \\
  N20-4-cw     & N20 & BSG         & 1.4 &            14.3 &            15.7 & 145 \\
  L15-1-cw     & L15 & RSG         & 1.6 &            13.7 &            15.3 & 146 \\
  W15-2-cw     & W15 & RSG         & 1.4 &            14.0 &            15.4 & 148 \\
  W15-2-cw-IIb & IIb & RSG He core & 1.5 & \hphantom{1}3.7 & \hphantom{1}5.1 & \hphantom{1}18 \\
 \enddata
 \tablenotetext{a}{Notation of \citet{wongwathanarat15}. The first
   letter does not correspond to any physical quantity (related to
   creators). The two-digit number is approximately the zero-age mass
   in M$_{\sun}$. The single-digit number indicates the model number
   in the series of models varying the explosion energy and initial
   seed perturbation~\citep{wongwathanarat13}. The last two letters
   are ``pw'' for power-law wind or ``cw'' for constant-wind
   boundary.}
 \tablenotetext{b}{At the time of the explosion.}
\end{deluxetable*}
\begin{deluxetable*}{lcccccc}
  \tablecaption{Ejecta Mass Compositions\label{tab:mas_com}}
  \tablewidth{0pt}
  \tablehead{
    \colhead{Model} & \colhead{$M(\mathrm{H})$} & \colhead{$M(\mathrm{He})$} & \colhead{$M(\mathrm{C})$} & \colhead{$M(\mathrm{O})$} & \colhead{$M(\mathrm{Ne})$} & \colhead{$M(\mathrm{Mg})$}\\
    \colhead{} & \colhead{(M$_{\sun}$)} & \colhead{(M$_{\sun}$)} & \colhead{(M$_{\sun}$)} & \colhead{(M$_{\sun}$)} & \colhead{(M$_{\sun}$)} & \colhead{(M$_{\sun}$)}} \startdata
  B15 &                $8.2$ &                 $5.4$ &  $1.2\times{}10^{-1}$ &                 0.2 & $0.5\times{}10^{-1}$ & $6.3\times{}10^{-3}$ \\
  N20 &                $5.7$ &                 $6.5$ &  $1.0\times{}10^{-1}$ &                 1.3 & $2.4\times{}10^{-1}$ & $1.7\times{}10^{-1}$ \\
  L15 &                $7.3$ &                 $5.3$ &  $1.8\times{}10^{-1}$ &                 0.6 & $1.0\times{}10^{-1}$ & $4.6\times{}10^{-2}$ \\
  W15 &                $7.2$ &                 $5.4$ &  $2.3\times{}10^{-1}$ &                 0.7 & $1.7\times{}10^{-1}$ & $5.0\times{}10^{-2}$ \\
  IIb &                $0.5$ &                 $2.1$ &  $1.7\times{}10^{-1}$ &                 0.6 & $1.3\times{}10^{-1}$ & $3.7\times{}10^{-2}$ \\\hline
      &     $M(\mathrm{Si})$ &       $M(\mathrm{S})$ &      $M(\mathrm{Ar})$ &     $M(\mathrm{Ca})$ &     $M(\mathrm{Fe})$ &                  Sum \\
      &         (M$_{\sun}$) &          (M$_{\sun}$) &          (M$_{\sun}$) &         (M$_{\sun}$) &         (M$_{\sun}$) &         (M$_{\sun}$) \\\hline
  B15 & $7.9\times{}10^{-2}$ &  $9.6\times{}10^{-3}$ &  $4.1\times{}10^{-3}$ & $2.1\times{}10^{-2}$ &                 0.10 &                 14.2 \\
  N20 & $8.1\times{}10^{-2}$ &  $9.3\times{}10^{-3}$ &  $4.7\times{}10^{-3}$ & $2.8\times{}10^{-2}$ &                 0.13 &                 14.3 \\
  L15 & $3.9\times{}10^{-2}$ &             \nodata{} &             \nodata{} & $4.5\times{}10^{-2}$ &                 0.15 &                 13.7 \\
  W15 & $4.3\times{}10^{-2}$ &             \nodata{} &             \nodata{} & $6.0\times{}10^{-2}$ &                 0.14 &                 14.0 \\
  IIb & $2.9\times{}10^{-2}$ &             \nodata{} &             \nodata{} & $5.7\times{}10^{-2}$ &                 0.13 &                  \hphantom{1}3.7 \\
  \enddata
\end{deluxetable*}
SN explosion models~\citep[][and references therein]{wongwathanarat15}
are used in order to estimate the X-ray absorption in SN
remnants. Details regarding the models are listed in
Table~\ref{tab:mod} and ejecta compositions in
Table~\ref{tab:mas_com}. We choose the B15 and N20 blue supergiant
(BSG) models because they are designed to match the progenitor
properties of \sna{}. Observations of \sna{} show that the progenitor
was a BSG~\citep{west87, white87, kirshner87, walborn87}. The B15
model is used for the analysis of X-ray limits on the \sna{} compact
object~\citep{alp18} because it is the single-star progenitor that
yields the best agreement of light-curve calculations based on
self-consistent 3D explosion models compared to observations of
\sna{}~\citep{utrobin15, utrobin18}.

The L15 and W15 models are included in the analysis with the purpose
of extending the results to a case considering an explosion of a red
supergiant (RSG), and comparing these results to the BSG cases, which
should be of interest for the X-ray analyses of more distant
SNe. Finally, the IIb model is an explosion of a bare helium core of a
RSG. The model shows similarities of the ejecta asymmetries to those
deduced from observations of \cas{}~\citep{wongwathanarat17}, and thus
is used for our \cas{} CCO analysis. We will refer to B15, N20, L15,
and W15 as the non-stripped models.

\subsection{Progenitors}\label{sec:pro}
All progenitor models are based on one-dimensional (1D) evolution of
non-rotating single stars. The B15 model represents SN 1987A and was
evolved without mass loss by \citet{woosley88}. The N20 model was
artificially constructed by combining the core~\citep{nomoto88} and
envelope~\citep{saio88} of two different stars. The combination was
made by \citet{shigeyama90} and designed to match the properties of
the progenitor of SN 1987A. The L15 model was computed without mass
loss by \citet{limongi00} and the W15 model is the s15s7b2 model of
\citet{woosley95}. The \iib{} model is created by removing all but
\henv{}\Msun{} of the hydrogen envelope of the W15 model to avoid
dynamical consequences of the hydrogen shell and to allow the star to
explode as a Type~IIb SN, which was the type of the Cas~A
SN~\citep{krause08, rest11}.

\subsection{Early Evolution}\label{sec:ear}
The short time between the start of core collapse to ${\sim}10$~ms
after bounce was treated by dedicated simulations. The B15 model was
collapsed using a Lagrangian hydro-code by \citet{bruenn93} and the
N20, L15, and W15 models were collapsed using the
\textsc{Prometheus-Vertex} code~\citep{rampp02} in one
dimension~\citep{wongwathanarat13}.

The simulation then were continued~\citep{wongwathanarat10,
  wongwathanarat13} from ${\sim}$10~ms after core bounce, past the
shock revival at ${\sim}250$~ms, and extended to about 1~s after shock
revival in three dimensions. These 3D simulations were performed using
the \textsc{Prometheus} code~\citep{fryxell91, muller91} with suitable
tuned neutrino heating at the inner boundary. When mapping from 1D to
3D, perturbations with an amplitude of 0.1\,\% were introduced by
hand~\citep{wongwathanarat13}. The 3D simulations were run on
axis-free Yin-Yang grids~\citep{kageyama04, wongwathanarat10b} with an
angular resolution of 2\degr{}. Because of computational limitations,
the innermost regions corresponding to radial velocities of less than
${\sim}10$\kmps{} were removed from the simulation~\citep{scheck06,
  wongwathanarat13}.

\subsection{Intermediate Evolution}\label{sec:int}
We let intermediate times refer to the time between the end of early
stages at ${\sim}1$~s and the beginning of late times at
${\sim}10^{5}$~s (${\sim}1$~day), which was followed by
\citet{wongwathanarat15}. The five models were mapped onto a larger
Yin-Yang grid with the same angular resolution of 2\degr{} and a
relative radial resolution higher than 1\,\% at all radii. The inner
boundary was initially set to 500~km and was continuously moved
outward to relax time step constraints imposed by the material at
small radii with high sound speed. The outer boundary was moved so
that the ejecta remain within the grid and was set to $10^{9}$~km by
the end of the simulations.

\subsection{Late Evolution}\label{sec:lat}
Shortly before the shocks break out from the progenitors
($t\lesssim10^4$~s), we continue the 3D simulations with a new version
of the \textsc{Prometheus-HOTB} code that includes a description of
$\beta$-decay. The grid remains unchanged compared to the intermediate
evolution. However, in this part of the simulation, we employ a
radially outwards moving grid, which reduces the numerical diffusion
between neighboring grid cells due to the lower relative velocities of
the radially expanding ejecta. We continue the simulations until
${\sim}150$ days for the non-stripped models and ${\sim}18$ days for
the \iib{} model. At later times, we do not expect significant effects
of the $\beta$\nobreakdash{-}decay on the structures, because the
ejecta become transparent to the released $\gamma$-radiation and
because most of the radioactive material has already decayed~(for
details, see Gabler et al.\ 2018, in preparation).

\added{
  \subsection{Asymmetries}\label{sec:met_asy}
  \begin{figure*}
    \centering
    \includegraphics[width=0.34\textwidth]{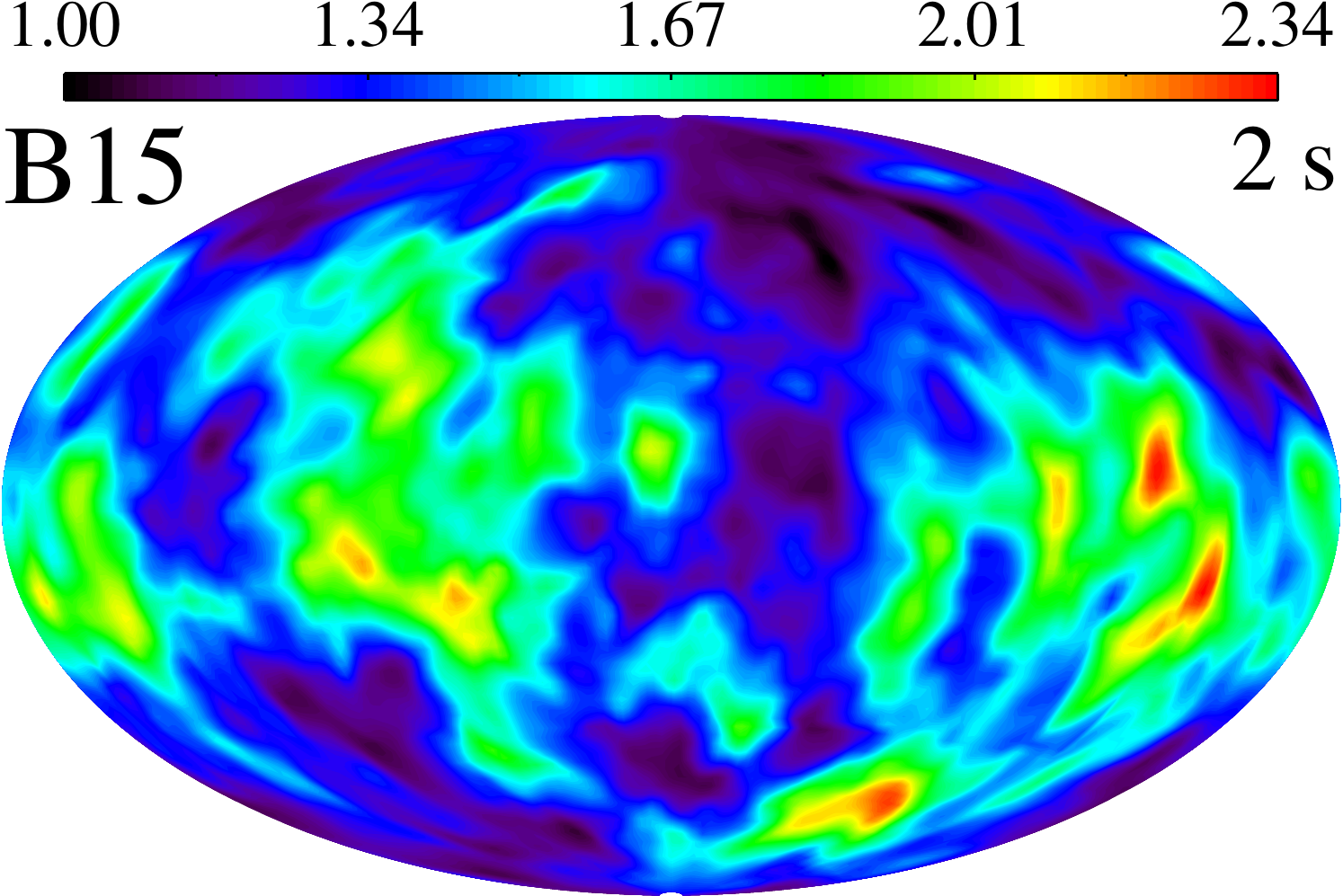}\hspace{48pt}
    \includegraphics[width=0.34\textwidth]{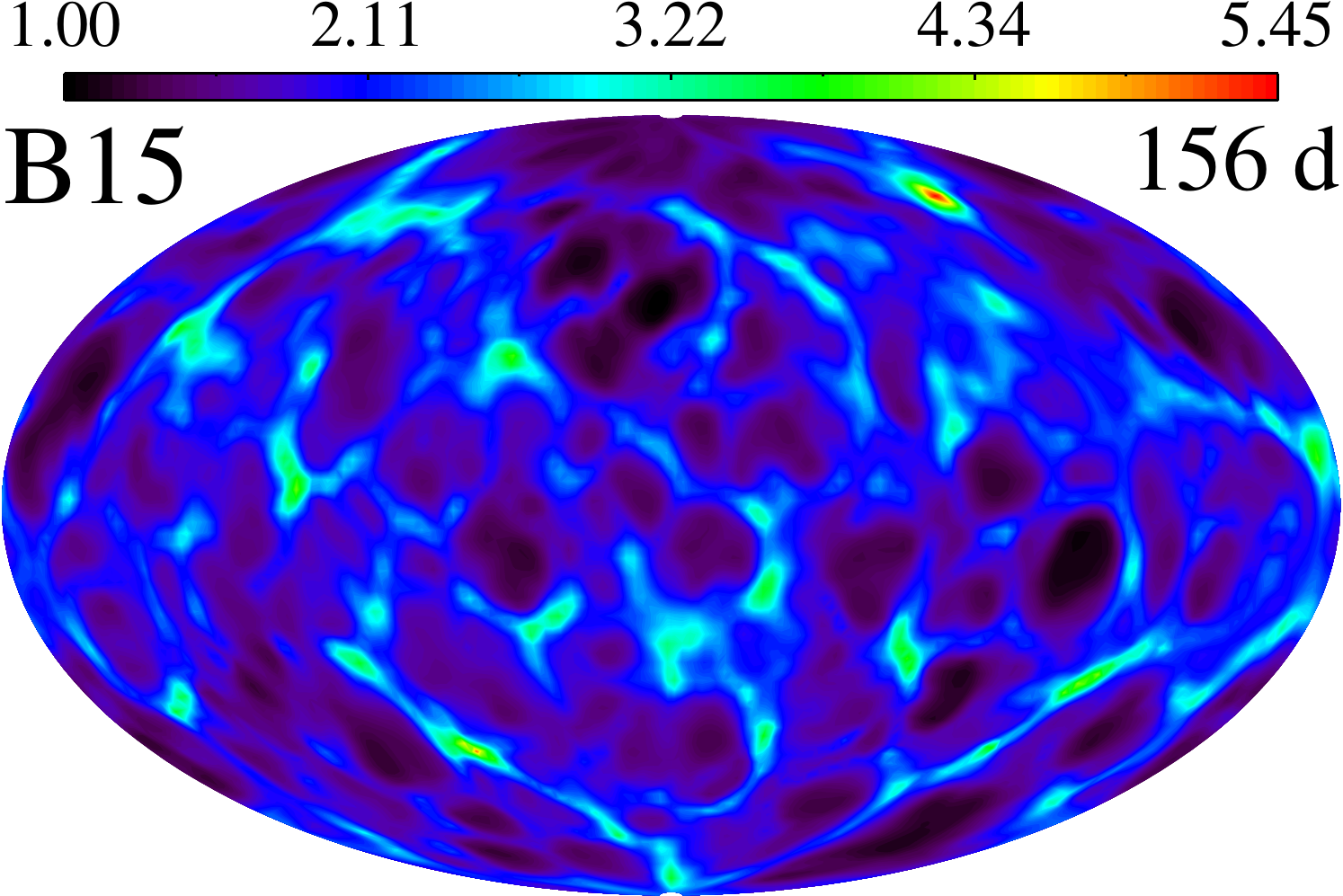}
    \includegraphics[width=0.34\textwidth]{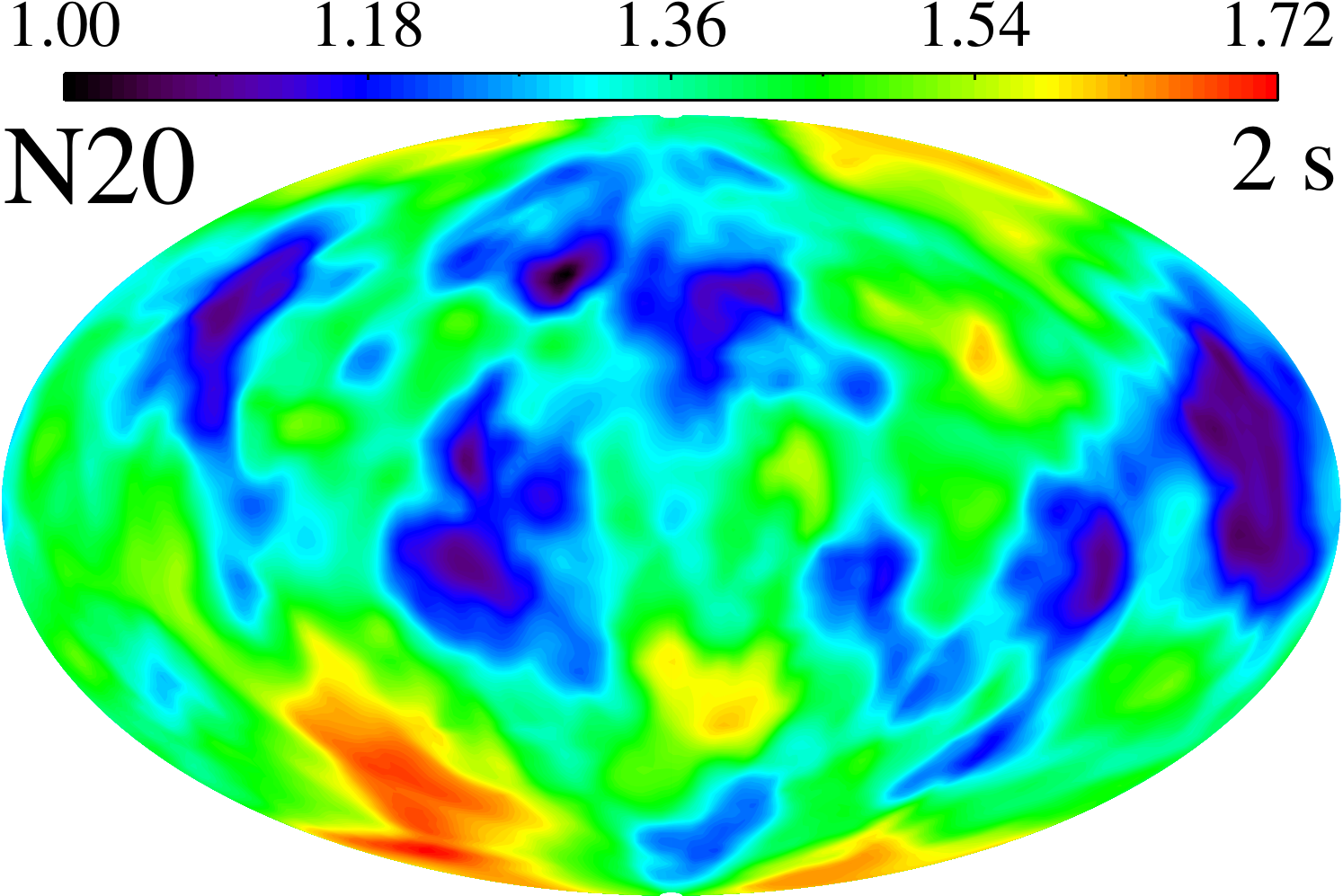}\hspace{48pt}
    \includegraphics[width=0.34\textwidth]{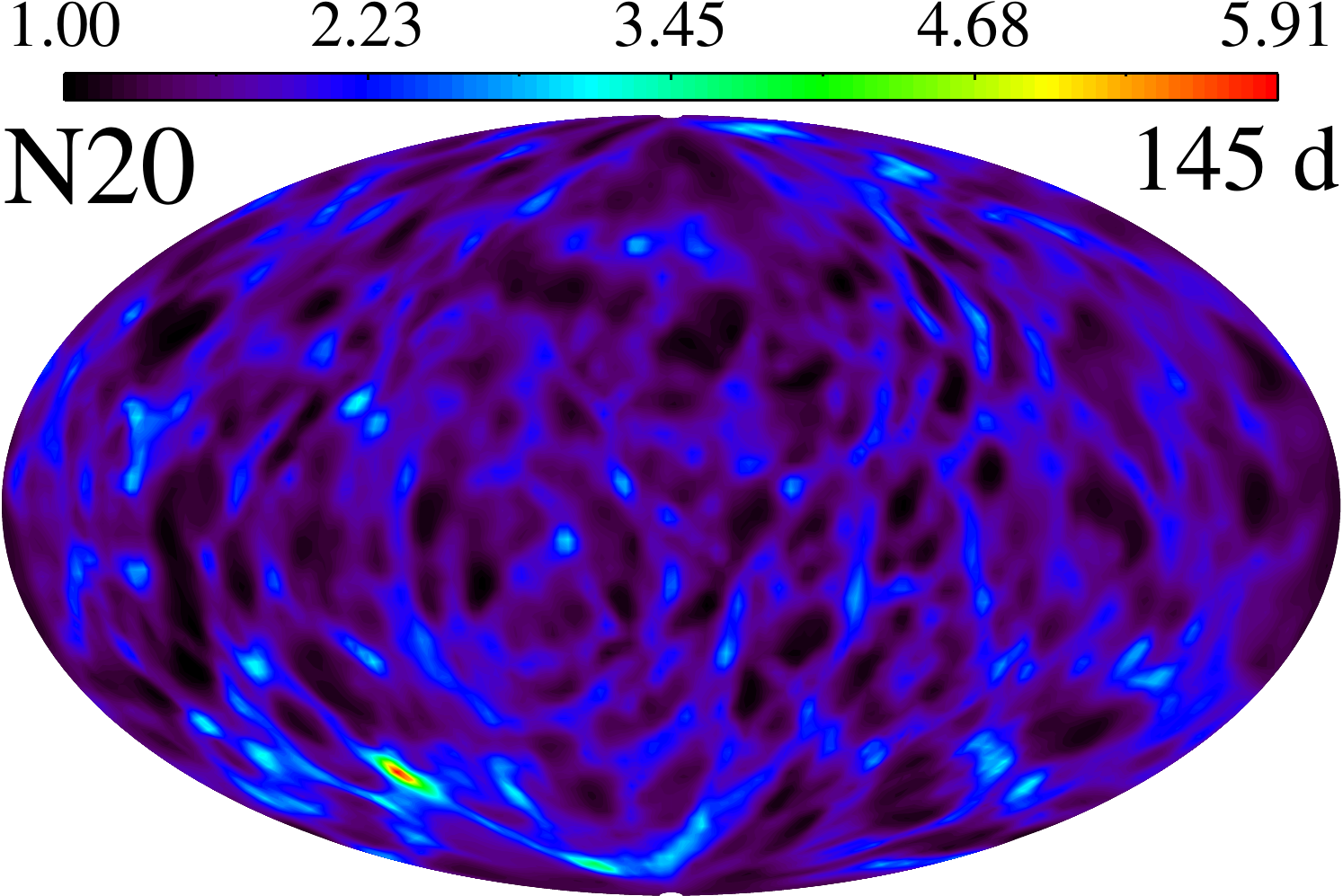}
    \includegraphics[width=0.34\textwidth]{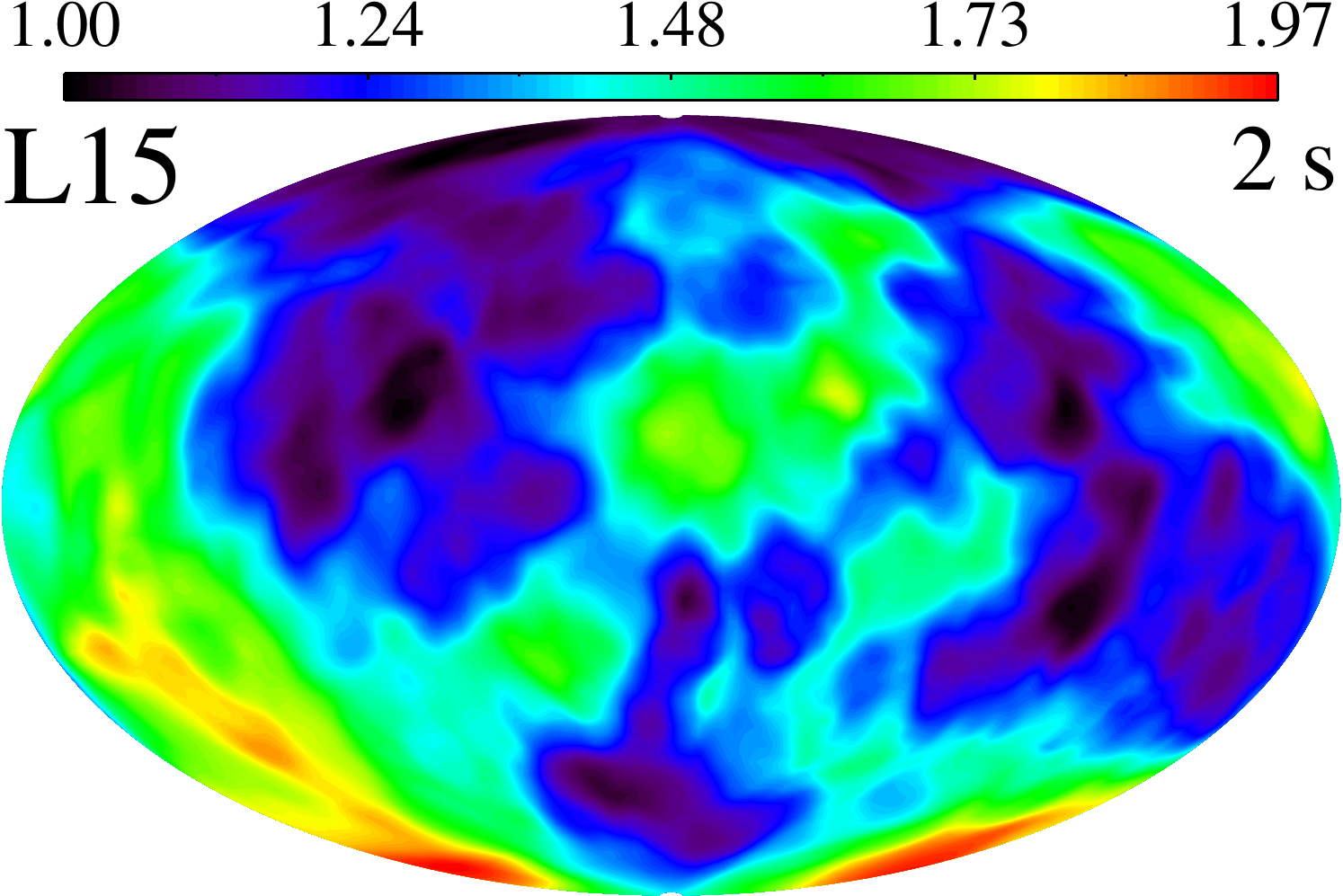}\hspace{48pt}
    \includegraphics[width=0.34\textwidth]{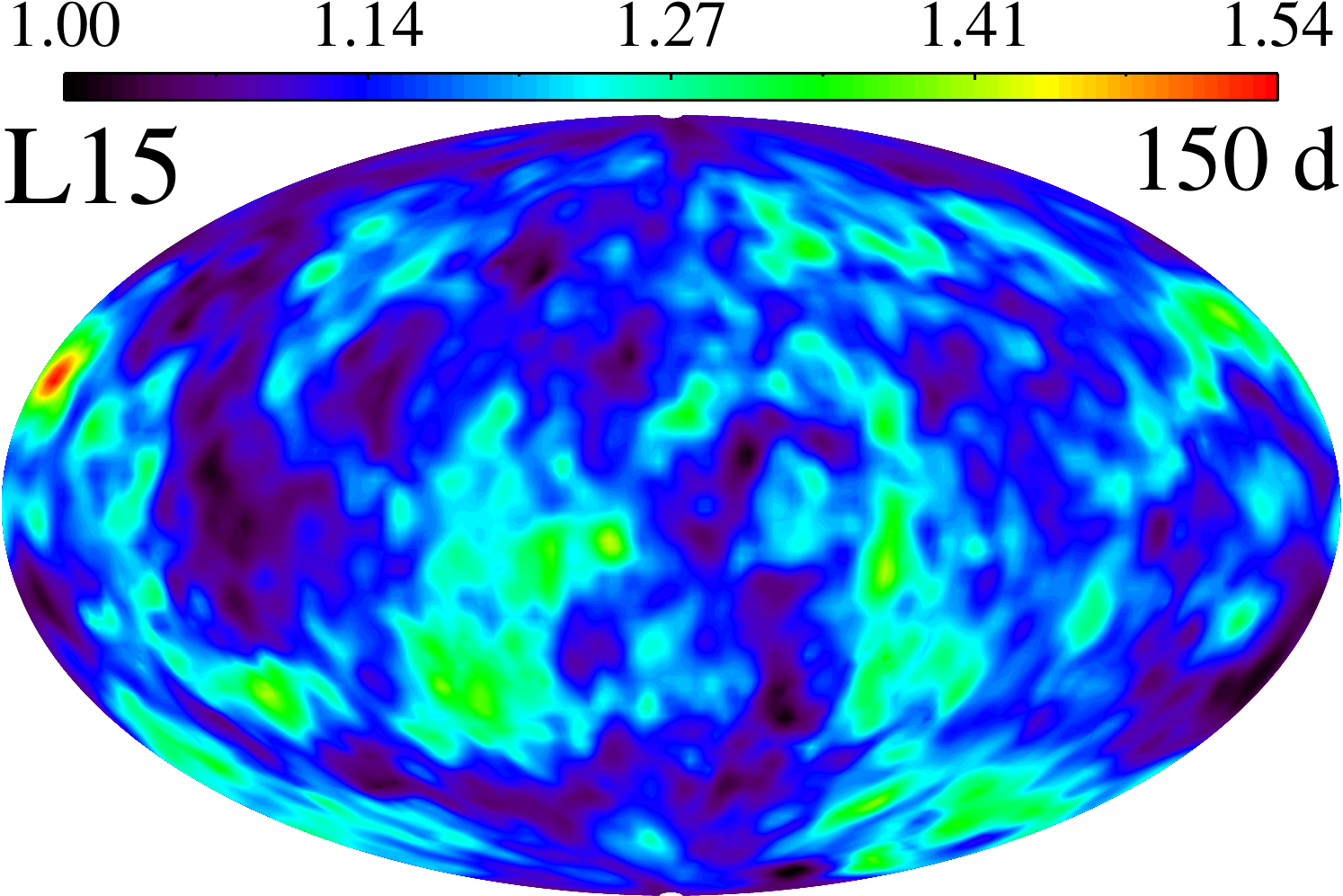}
    \includegraphics[width=0.34\textwidth]{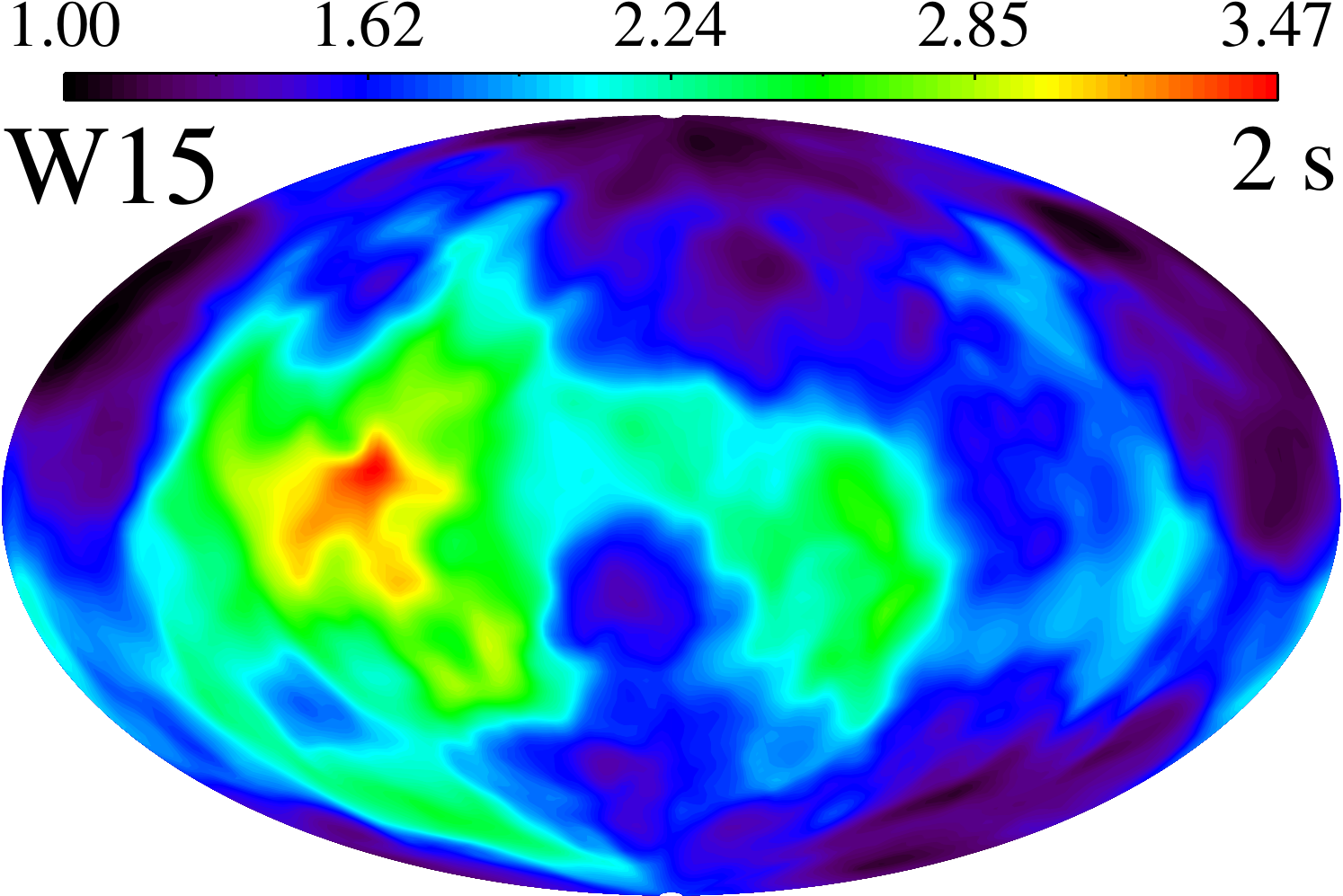}\hspace{48pt}
    \includegraphics[width=0.34\textwidth]{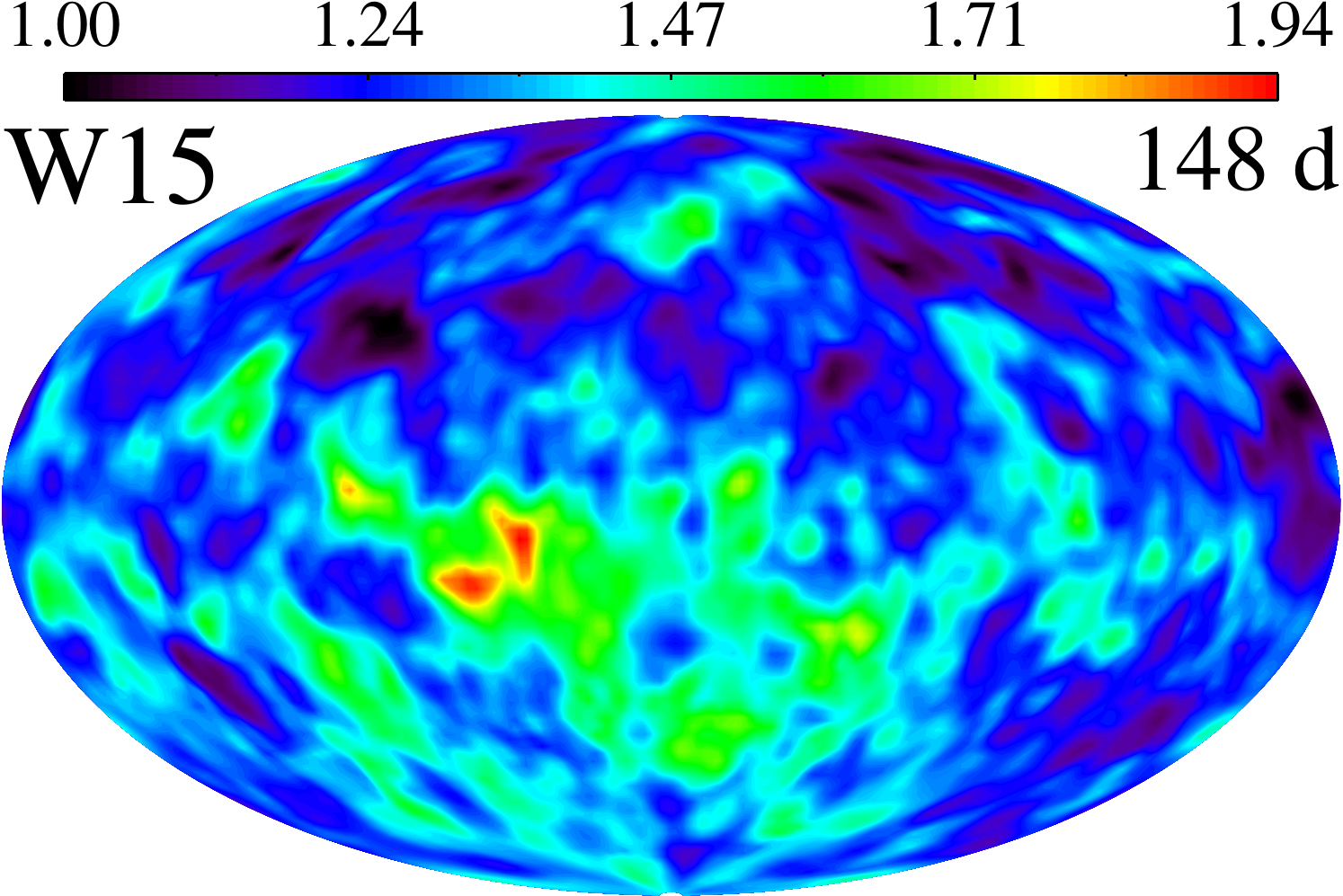}
    \includegraphics[width=0.34\textwidth]{W15_intrho_early-crop.pdf}\hspace{48pt}
    \includegraphics[width=0.34\textwidth]{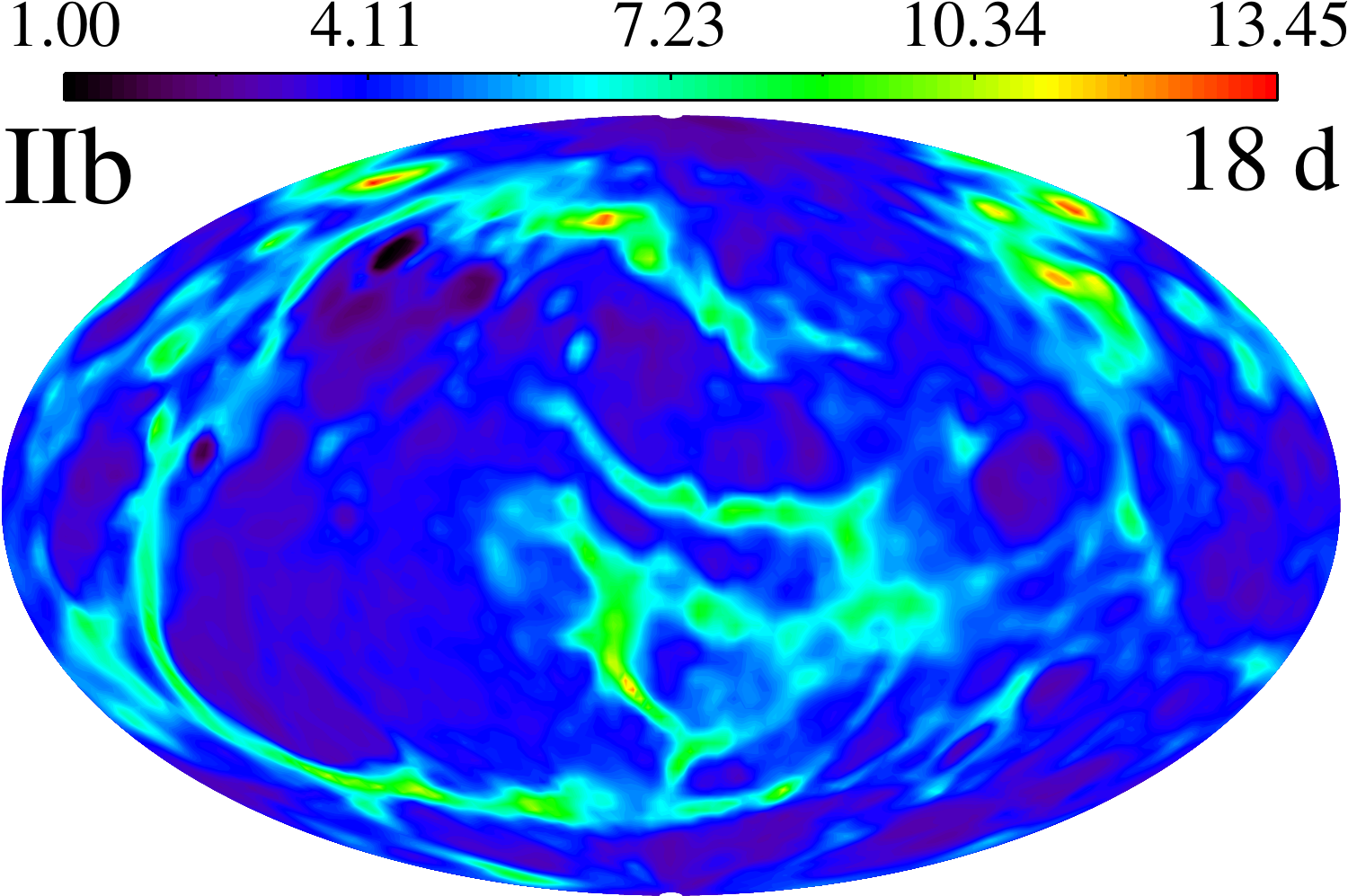}
    \caption{Equal-area Hammer projections of the column mass
      densities at 2~s (left) and at $t_\text{late}$ (right). The
      colors are normalized to the minimum column density for each
      model. The bottom-left projection is labeled W15 because the
      \iib{} model is created from the W15 model. Larger structures
      are present at early times and during the evolution the
      large-scale structures fragment into smaller filaments and
      fingers. We note that the contrast between different directions
      is much larger for the IIb model.\label{fig:mod_asy}}
  \end{figure*}
  All models possess strong asymmetries that are created as a result
  of non-radial hydrodynamic instabilities behind the stalled shock,
  where neutrinos deposit the energy of the supernova explosion during
  the first seconds after core bounce~\citep{wongwathanarat15}. This
  innermost region is dominated by iron-peak elements and form
  fast-rising plumes, which are initially rather similar for all
  models. In the carbon-oxygen layer, the plumes are not extremely
  decelerated in all models except for the N20 case, which results in
  considerably more compressed structures in the N20 model. More
  pronounced differences between the models develop at the interface
  between the carbon-oxygen core and the helium layer, where
  Rayleigh-Taylor instabilities after the passage of the shock change
  the morphology of the expanding ejecta. Considerable evolution of
  the morphology also occurs after the shock has decelerated in the
  hydrogen envelope. A strong reverse shock is launched upstream into
  the expanding inner ejecta, which decelerates and compresses the
  ejecta. The \iib{} model evolves very similarly to the W15 model
  except that no reverse shock is formed at the helium-hydrogen
  interface. This results in stronger and larger-scale asymmetries,
  higher overall ejecta velocities, and a sharper inner radius of the
  ejecta (because of the absence of an extended low-velocity tail of
  the radial distribution) in the \iib{} model. During the time
  covered until the final stage of our simulations, the $\beta$-decay
  inflates the radioactive iron-peak fingers significantly. A
  comparison of the asymmetries at 2~s and $t_\mathrm{late}$ between
  all considered models is provided in Fig.~\ref{fig:mod_asy}. The
  plots also visualize the outcome of different mixing and
  fragmentation histories of the initial structures as a consequence
  of differences in the evolutionary phases mentioned before.

  Additionally, it has been shown that different asymmetries develop
  even if the initial conditions are nearly identical for the same
  progenitor because of the chaotic nature of hydrodynamic
  instabilities~\citep{wongwathanarat15}. \citet{muller17}
  demonstrated by 3D simulations that pre-collapse large-scale
  convective perturbations in the oxygen shell during the late stages
  of stellar evolution can have a supportive influence on the onset of
  the explosion. These seed perturbations are important because they
  help the development of violent hydrodynamic mass motions behind the
  supernova shock, which facilitate the outward expansion of the
  supernova shock and of the neutrino-heated matter in its wake.  More
  details on the evolution of asymmetries can be found in Section~5 of
  \citet{wongwathanarat15} for the early to intermediate phases, in
  M.~Gabler et al.~(2018, in preparation) for the late phases of the
  non-stripped models, and in Section~3.3 of \citet{wongwathanarat17}
  for late stages of the \iib{} model, which exemplifies the
  consequences of removing the hydrogen envelope from the W15 model.
}

\subsection{Nucleosynthesis}\label{sec:nuc}
There were minor differences in the treatment of the nucleosynthesis
in the different models. Starting from after core bounce at
${\sim}10$~ms, the nucleosynthesis in the BSG models was followed by a
network of elements that includes protons; the 13
$\alpha$\nobreakdash{-}nuclei from $^{4}$He to $^{56}$Ni; radioactive
daughter products; and a tracer nucleus X~\citep{kifonidis03,
  wongwathanarat13, wongwathanarat15}. The tracer comprises
neutron-rich, iron-group elements and was produced in grid cells where
the electron fraction was below 0.49. The radioactive products that
were included were $^{44}$Ca and $^{44}$Sc from $\beta^{+}$ decay of
$^{44}$Ti, and $^{56}$Fe and $^{56}$Co from $\beta^{+}$ decay of
$^{56}$Ni. The RSG and \iib{} models follow fewer nuclear species,
omitting $^{32}$S, $^{36}$Ar, $^{48}$Cr, and $^{52}$Fe in the burning
network. Free neutrons are followed in all simulations but are
irrelevant to our study of photoabsorption and Compton scattering.

\section{Methods}\label{sec:methods}
The models consist of 3D maps of number densities of each element,
defined from an inner radius to an outer radius. The vast majority of
the mass of the SN ejecta is inside the spherical shell defined by the
simulation grid. We refer to this mass as the ejecta mass
$M_\mathrm{ej}$. The outer radius is set such that essentially all of
the ejecta is contained within the grid. Dynamics of the material
inside of the inner boundary are not followed in the 3D explosion
simulations. This material includes the NS with a baryonic mass
$M_\mathrm{NS}$ and its surrounding material. The mass that is
excluded from the simulation within the inner radius has velocities
below escape velocity at the end of our simulations and is likely to
fall back to the NS on a longer timescale. It is therefore added to
the mass of the compact remnant. Typical masses are 0.01\Msun{} for
the non-stripped models and less than 0.15\Msun{} for the \iib{}
model. We verified that any realistic contribution from the mass
inside the inner cut-off has a negligible impact on the results even
if the material escaped.

The data are mapped from the Yin-Yang grids used for the 3D explosion
simulations to spherical coordinate systems for the current work. We
denote the conventional radial coordinate $r$ but define $\phi$ as the
longitude ranging from 0 to $2\pi$ and $\theta{}$ as the latitude
ranging from $-\pi/2$ to $\pi/2$. The resolution $\Delta r/r$ is
better than 1\,\% at all radii except within radii corresponding to
velocities $v < 100$\kmps{}, where the resolution is better than
10\,\%. The angular resolution of $2\degr{}$ is uniform in $\phi$ and
$\theta$. These spherical grids resolve the models well enough for
accurate estimates of the optical depth while not over-resolving the
Yin-Yang grids used for the 3D explosion simulations.

We denote a unit vector starting from the origin
$\boldsymbol{\hat{n}} = (1, \phi{}, \theta{})$, which will henceforth
be referred to as ``direction''. Contributions from different
directions are weighted by $\cos\theta$ throughout the analysis to
compensate for the varying angular resolution of the spherical
coordinate system. The expansions of all models are assumed to be
homologous after $t_\text{late}$ and all presented values are scaled
to 10\,000~days (${\sim}27$~years). The column number density as a
function of direction $N_\mathrm{SN}(\boldsymbol{\hat{n}})$ is
computed by integrating the column number density along all directions
of the model specified by the angular coordinates $\phi$ and
$\theta$. The integration is carried out separately for each chemical
element. The optical depth $\tau$ can then be computed using
\begin{equation}
  \label{eq:def_tau}
  \tau(E, \boldsymbol{\hat{n}}) = \sum_i [\sigma_\mathrm{SN}(i, E)+\sigma_\mathrm{C}(i, E)]\, N_\mathrm{SN}(i, \boldsymbol{\hat{n}}),
\end{equation}
where $E$ is the photon energy, the summation index $i$ represents all
chemical elements, $\sigma_\mathrm{SN}$ is the photoabsorption
cross-section, and $\sigma_\mathrm{C}$ is the Compton scattering
cross-section. The photoabsorption cross-sections below 10~keV are
taken from the tables of \citet{gatuzz15}\footnote{These are also the
  tables used by the XSPEC model
  \texttt{ISMabs}~\citep{gatuzz15}.}. They compile cross-sections
obtained using different analytic methods~\citep{bethe57}, the
$R$-matrix method~\citep{gorczyca00, garcia05, garcia09, juett06,
  witthoeft09, witthoeft11, gorczyca13, hasoglu14}, semi-empirical
fits~\citep{verner95}, and experimental
values~\citep{kortright00}. The tables cover energies from 0.01~keV to
10~keV with 650\,000 logarithmically spaced steps, which results in a
relative energy resolution of $10^{-5}$. Photoabsorption
cross-sections from 10 to 100~keV are computed using the analytic fits
of \citet{verner95}. The Compton scattering cross-section at energies
higher than ${\sim}$10~keV, which is the binding energies of the
K-shell electrons, is given by the Klein-Nishina
formula~\citep[e.g.][]{rybicki79}.

We reduce the number of free parameters by making a number of
simplifications. We assume that the formation of molecules and dust
has a negligible impact on the X-ray absorption properties (see
Section~\ref{sec:errors}). It is also assumed that all atoms are in
the neutral ground state. Cross-sections for single and double
ionization are included in Table~2 of \citet{gatuzz15} and general
cases can be explored using the analytic fits of \citet{verner95} and
\citet{verner96}. Detailed models of \sna{} show that the ejecta
indeed is mainly in neutral and at most in singly ionized
states~\citep[low ionization ions such as Mg\textsc{\,ii},
Si\textsc{\,ii}, Ca\textsc{\,II}, and
Fe\textsc{\,ii},][]{jerkstrand11}. This is also expected for young SNe
in general. For remnants like Cas A, the situation is more
complicated. The effect of ionization above 0.3~keV is small in most
cases, but quantitative estimates would require detailed knowledge of
the ionization state of the gas.

Another simplification is that we group some of the elements traced by
the nucleosynthesis network. This is done because of radioactive decay
and because we aim to provide column densities of elements that are
abundant and included in publicly available ISM models. We combine the
number densities of $^{40}$Ca, $^{44}$Ca, $^{44}$Sc, $^{44}$Ti, and
$^{48}$Cr and treat the combined number density as $^{40}$Ca. The
elements $^{44}$Ti, $^{44}$Sc, and $^{44}$Ca form a radioactive decay
chain with $^{44}$Ca as the stable product. The difference between
isotopes of the same element is negligible because the cross-sections
are determined by the electrons. For reference, the number of neutrons
of an atom is not a parameter in the analytic fits of the
photoabsorption cross-sections~\citep{verner95, verner96} and the
Klein-Nishina formula. The element $^{48}$Cr is treated as $^{40}$Ca
because $^{48}$Cr decays into $^{48}$Ti, which makes Ca the most
similar element commonly included in absorption models in XSPEC. We
apply the same grouping to $^{52}$Fe, $^{56}$Fe, $^{56}$Co, $^{56}$Ni,
and X and treat the group as $^{56}$Fe. The elements $^{56}$Ni,
$^{56}$Co, and $^{56}$Fe constitute the decay chain from $^{56}$Ni to
the stable isotope $^{56}$Fe. At the epochs of interest for \sna{} and
\cas{} the abundant isotopes, $^{56}$Ni and $^{56}$Co have all decayed
into $^{56}$Fe. The mass numbers of all atoms will henceforth be
omitted.

\section{Results}\label{sec:results}
A complete absorption description requires knowledge of the column
densities of all elements for all directions. We do not intend to
provide a description at this level of detail. Instead, we focus on
direction-averaged column number densities and optical depths in
Section~\ref{sec:dir_avg}, with the purpose of providing
representative values for quantities of interest. The optical depth
variations along different directions are explored in
Section~\ref{sec:asy}. This represents the level of variance of the
optical depth within a specific SNR. We superficially investigate the
differences introduced by different compositions along different lines
of sight in Section~\ref{sec:eje_com}. Results concerning Compton
scattering are contained in Section~\ref{sec:com_sca}. A comparison of
SN ejecta absorption with ISM absorption is included in
Appendix~\ref{app:com}. Throughout the analysis, we assume that atoms
are in the neutral ground state and that the formation of molecules
and dust has a negligible impact on the relevant absorption properties
(Section~\ref{sec:methods}).

\added{All presented results are at a time of 10\,000 days unless
  otherwise stated. However, the column densities and optical depths
  can be scaled to other epochs as $\propto t^{-2}$, where $t$ is the
  time since the SN explosion, if the ejecta are assumed to expand
  homologously. The models are evolved up to the point when the ejecta
  start expanding homologously, implying that $t_\text{late}$ roughly
  marks the start of the homologous phase. Homologous expansion is a
  reasonable approximation until the reverse shock reaches the
  material with velocities of ${\sim}10^3$\kmps{}, which dominates the
  contribution to the optical depths (Section~\ref{sec:dir_avg}). The
  time at which this happens varies over more than an order of
  magnitude because the dynamics are sensitive to explosion energy,
  ejecta mass, and density distribution of the ejecta and
  circumstellar medium along the line of sight. For \sna{}, the inner
  ejecta along our line of sight can safely be assumed to be expanding
  freely at present epochs for our purposes. For \cas{}, the reverse
  shock has reached ejecta expanding at
  ${\sim}$5500\kmps{}~\citep{morse04}. In general, theoretical models
  predict that the inner ejecta expand freely for up to a few thousand
  years for simple models of core-collapse
  SNe~\citep[e.g.][]{truelove99}. However, in extreme cases, such as
  for Type~IIn SNe and for lines of sight crossing the dense inner
  ring of \sna{}~\citep{fransson13}, the timescales can be much
  shorter.}

\subsection{Direction-Averaged Optical Depths}\label{sec:dir_avg}
\begin{deluxetable*}{lcccclccccc}
  \tablecaption{Direction-Averaged Column Number Densities at 10\,000 days\label{tab:abundances}}
  \tablewidth{0pt}
  \tablehead{
    \colhead{Elem.} & \colhead{$\langle N_\mathrm{SN}\rangle_{\boldsymbol{\hat{n}}}$\tablenotemark{a}} & \colhead{$\langle A_\mathrm{SN}\rangle_{\boldsymbol{\hat{n}}}$\tablenotemark{b}} &
    \colhead{$\langle A_\mathrm{SN}\rangle_{\boldsymbol{\hat{n}}}$\tablenotemark{c}} & \colhead{$\langle N^\mathrm{e}_\mathrm{SN}\rangle_{\boldsymbol{\hat{n}}}$\tablenotemark{d}} & \colhead{Elem.} & \colhead{$\langle N_\mathrm{SN}\rangle_{\boldsymbol{\hat{n}}}$} &
    \colhead{$\langle A_\mathrm{SN}\rangle_{\boldsymbol{\hat{n}}}$} & \colhead{$\langle A_\mathrm{SN}\rangle_{\boldsymbol{\hat{n}}}$} &\colhead{$\langle N^\mathrm{e}_\mathrm{SN}\rangle_{\boldsymbol{\hat{n}}}$} \\
    \colhead{} & \colhead{(cm$^{-2}$)} & \colhead{} & \colhead{(A$_\mathrm{ISM}$)} & \colhead{(cm$^{-2}$)} & \colhead{} & \colhead{(cm$^{-2}$)} & \colhead{} & \colhead{(A$_\mathrm{ISM}$)} &\colhead{(cm$^{-2}$)}} \startdata
                           \multicolumn{5}{c}{B15}                                 &                       \multicolumn{5}{c}{N20}                       \\\hline
  H   & $1.5\times{}10^{22}$ & $                 1$ &     $  1$ & $1.5\times{}10^{22}$ & H   & $9.5\times{}10^{21}$ & $                 1$ &    $   1$ & $9.5\times{}10^{21}$ \\
  He  & $1.2\times{}10^{22}$ & $7.8\times{}10^{-1}$ &     $  8$ & $2.4\times{}10^{22}$ & He  & $5.4\times{}10^{21}$ & $5.7\times{}10^{-1}$ &    $   6$ & $1.1\times{}10^{22}$ \\
  C   & $1.5\times{}10^{20}$ & $1.0\times{}10^{-2}$ &     $ 43$ & $9.0\times{}10^{20}$ & C   & $5.8\times{}10^{19}$ & $6.1\times{}10^{-3}$ &    $  26$ & $3.5\times{}10^{20}$ \\
  O   & $2.1\times{}10^{20}$ & $1.4\times{}10^{-2}$ &     $ 29$ & $1.7\times{}10^{21}$ & O   & $8.3\times{}10^{20}$ & $8.7\times{}10^{-2}$ &    $ 179$ & $6.6\times{}10^{21}$ \\
  Ne  & $4.5\times{}10^{19}$ & $3.0\times{}10^{-3}$ &     $ 34$ & $4.5\times{}10^{20}$ & Ne  & $9.8\times{}10^{19}$ & $1.0\times{}10^{-2}$ &    $ 119$ & $9.8\times{}10^{20}$ \\
  Mg  & $4.8\times{}10^{18}$ & $3.2\times{}10^{-4}$ &     $ 13$ & $5.8\times{}10^{19}$ & Mg  & $7.5\times{}10^{19}$ & $7.9\times{}10^{-3}$ &    $ 315$ & $9.0\times{}10^{20}$ \\
  Si  & $5.8\times{}10^{19}$ & $3.9\times{}10^{-3}$ &     $208$ & $8.1\times{}10^{20}$ & Si  & $3.8\times{}10^{19}$ & $4.1\times{}10^{-3}$ &    $ 218$ & $5.3\times{}10^{20}$ \\
  S   & $7.2\times{}10^{18}$ & $4.8\times{}10^{-4}$ &     $ 39$ & $1.2\times{}10^{20}$ & S   & $1.1\times{}10^{19}$ & $1.2\times{}10^{-3}$ &    $  95$ & $1.8\times{}10^{20}$ \\
  Ar  & $2.9\times{}10^{18}$ & $1.9\times{}10^{-4}$ &     $ 74$ & $5.2\times{}10^{19}$ & Ar  & $3.9\times{}10^{18}$ & $4.2\times{}10^{-4}$ &    $ 162$ & $7.0\times{}10^{19}$ \\
  Ca  & $1.4\times{}10^{19}$ & $9.4\times{}10^{-4}$ &     $591$ & $2.8\times{}10^{20}$ & Ca  & $2.4\times{}10^{19}$ & $2.6\times{}10^{-3}$ &    $1615$ & $4.8\times{}10^{20}$ \\
  Fe  & $6.0\times{}10^{19}$ & $4.0\times{}10^{-3}$ &     $148$ & $1.6\times{}10^{21}$ & Fe  & $5.9\times{}10^{19}$ & $6.3\times{}10^{-3}$ &    $ 232$ & $1.5\times{}10^{21}$ \\
  Sum & $2.7\times{}10^{22}$ &         \nodata{}    & \nodata{} & $4.4\times{}10^{22}$ & Sum & $1.6\times{}10^{22}$ &         \nodata{}    & \nodata{} & $3.2\times{}10^{22}$ \\\hline       
                                                     \multicolumn{5}{c}{L15}           &                                     \multicolumn{5}{c}{W15}                          \\\hline
  H   & $1.7\times{}10^{22}$ & $                 1$ &     $  1$ & $1.7\times{}10^{22}$ & H   & $2.1\times{}10^{22}$ & $                 1$ &    $   1$ & $2.1\times{}10^{22}$ \\
  He  & $5.9\times{}10^{21}$ & $3.5\times{}10^{-1}$ &     $  4$ & $1.2\times{}10^{22}$ & He  & $6.4\times{}10^{21}$ & $3.0\times{}10^{-1}$ &    $   3$ & $1.3\times{}10^{22}$ \\
  C   & $1.2\times{}10^{20}$ & $6.8\times{}10^{-3}$ &     $ 29$ & $7.2\times{}10^{20}$ & C   & $1.7\times{}10^{20}$ & $8.0\times{}10^{-3}$ &    $  33$ & $1.0\times{}10^{21}$ \\
  O   & $3.1\times{}10^{20}$ & $1.8\times{}10^{-2}$ &     $ 37$ & $2.5\times{}10^{21}$ & O   & $5.3\times{}10^{20}$ & $2.5\times{}10^{-2}$ &    $  51$ & $4.2\times{}10^{21}$ \\
  Ne  & $4.3\times{}10^{19}$ & $2.5\times{}10^{-3}$ &     $ 29$ & $4.3\times{}10^{20}$ & Ne  & $9.8\times{}10^{19}$ & $4.6\times{}10^{-3}$ &    $  53$ & $9.8\times{}10^{20}$ \\
  Mg  & $2.3\times{}10^{19}$ & $1.3\times{}10^{-3}$ &     $ 54$ & $2.8\times{}10^{20}$ & Mg  & $2.9\times{}10^{19}$ & $1.3\times{}10^{-3}$ &    $  53$ & $3.5\times{}10^{20}$ \\
  Si  & $2.1\times{}10^{19}$ & $1.2\times{}10^{-3}$ &     $ 67$ & $2.9\times{}10^{20}$ & Si  & $2.6\times{}10^{19}$ & $1.2\times{}10^{-3}$ &    $  65$ & $3.6\times{}10^{20}$ \\
  Ca  & $2.7\times{}10^{19}$ & $1.6\times{}10^{-3}$ &     $998$ & $5.4\times{}10^{20}$ & Ca  & $5.8\times{}10^{19}$ & $2.7\times{}10^{-3}$ &    $1705$ & $1.2\times{}10^{21}$ \\
  Fe  & $5.8\times{}10^{19}$ & $3.4\times{}10^{-3}$ &     $125$ & $1.5\times{}10^{21}$ & Fe  & $7.1\times{}10^{19}$ & $3.3\times{}10^{-3}$ &    $ 124$ & $1.8\times{}10^{21}$ \\
  Sum & $2.4\times{}10^{22}$ &         \nodata{}    & \nodata{} & $3.5\times{}10^{22}$ & Sum & $2.9\times{}10^{22}$ &         \nodata{}    & \nodata{} & $4.4\times{}10^{22}$ \\\hline
                                                                    \multicolumn{10}{c}{\iib{}}                                                                               \\\hline
  H   & $1.2\times{}10^{20}$ & $                 1$ &    $   1$ & $1.2\times{}10^{20}$ & Mg  & $1.7\times{}10^{18}$ & $1.4\times{}10^{-2}$ &    $ 545$ & $2.0\times{}10^{19}$ \\
  He  & $2.2\times{}10^{20}$ & $               1.8$ &    $  18$ & $4.4\times{}10^{20}$ & Si  & $1.2\times{}10^{18}$ & $9.3\times{}10^{-3}$ &    $ 502$ & $1.7\times{}10^{19}$ \\
  C   & $1.2\times{}10^{19}$ & $1.0\times{}10^{-1}$ &    $ 403$ & $7.2\times{}10^{19}$ & Ca  & $1.8\times{}10^{18}$ & $1.5\times{}10^{-2}$ &    $9444$ & $3.6\times{}10^{19}$ \\
  O   & $3.7\times{}10^{19}$ & $3.0\times{}10^{-1}$ &    $ 608$ & $3.0\times{}10^{20}$ & Fe  & $2.4\times{}10^{18}$ & $1.9\times{}10^{-2}$ &    $ 717$ & $6.2\times{}10^{19}$ \\
  Ne  & $6.4\times{}10^{18}$ & $5.2\times{}10^{-2}$ &    $ 592$ & $6.4\times{}10^{19}$ & Sum & $4.0\times{}10^{20}$ &         \nodata{}    & \nodata{} & $1.1\times{}10^{21}$ \\
  \enddata
  \tablenotetext{a}{Direction-averaged column number density of the SN
    ejecta.}
  
  \tablenotetext{b}{Direction-averaged SN abundance relative to
    hydrogen.}
  
  \tablenotetext{c}{Relative SN ejecta abundances in units of
    corresponding ISM abundances~\citep{wilms00}, equivalent to
    $\langle
    A_\mathrm{SN}\rangle_{\boldsymbol{\hat{n}}}/A_\mathrm{ISM}$.}
  
  \tablenotetext{d}{Direction-averaged column number density of
    electrons of the SN ejecta. Same as
    $\langle N_\mathrm{SN}\rangle_{\boldsymbol{\hat{n}}}$ scaled by
    the number of electrons of the corresponding chemical element.}
\end{deluxetable*}
The direction-averaged column densities of all elements at 10\,000
days are provided in Table~\ref{tab:abundances}. The SN abundances
relative to hydrogen are given by
$\langle A_\mathrm{SN}\rangle_{\boldsymbol{\hat{n}}}$ and
$A_\mathrm{ISM}$ are the ISM abundances~\citep{wilms00} normalized to
hydrogen. The SN ejecta have very high metallicities compared to the
ISM (see Appendix~\ref{app:com} for comparisons with ISM absorption).

\begin{figure*}
  \includegraphics[width=\columnwidth]{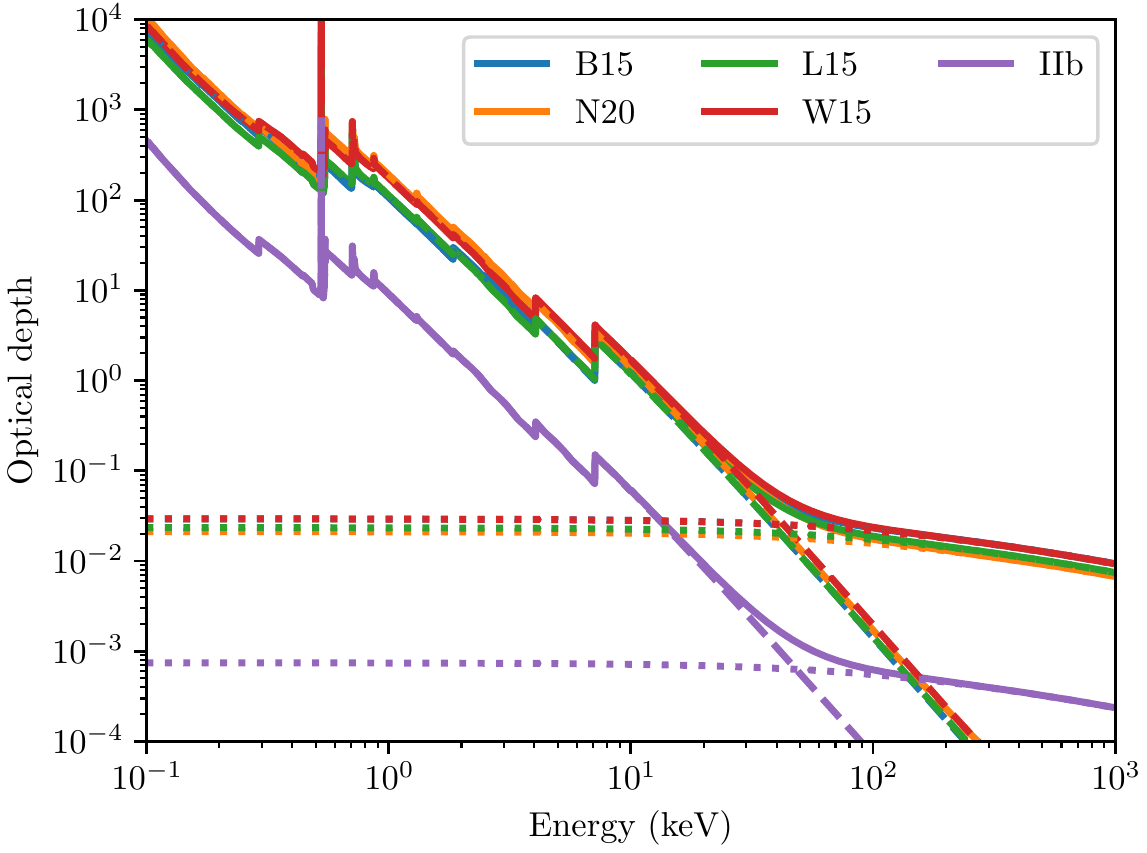}
  \hfill{}
  \includegraphics[width=\columnwidth]{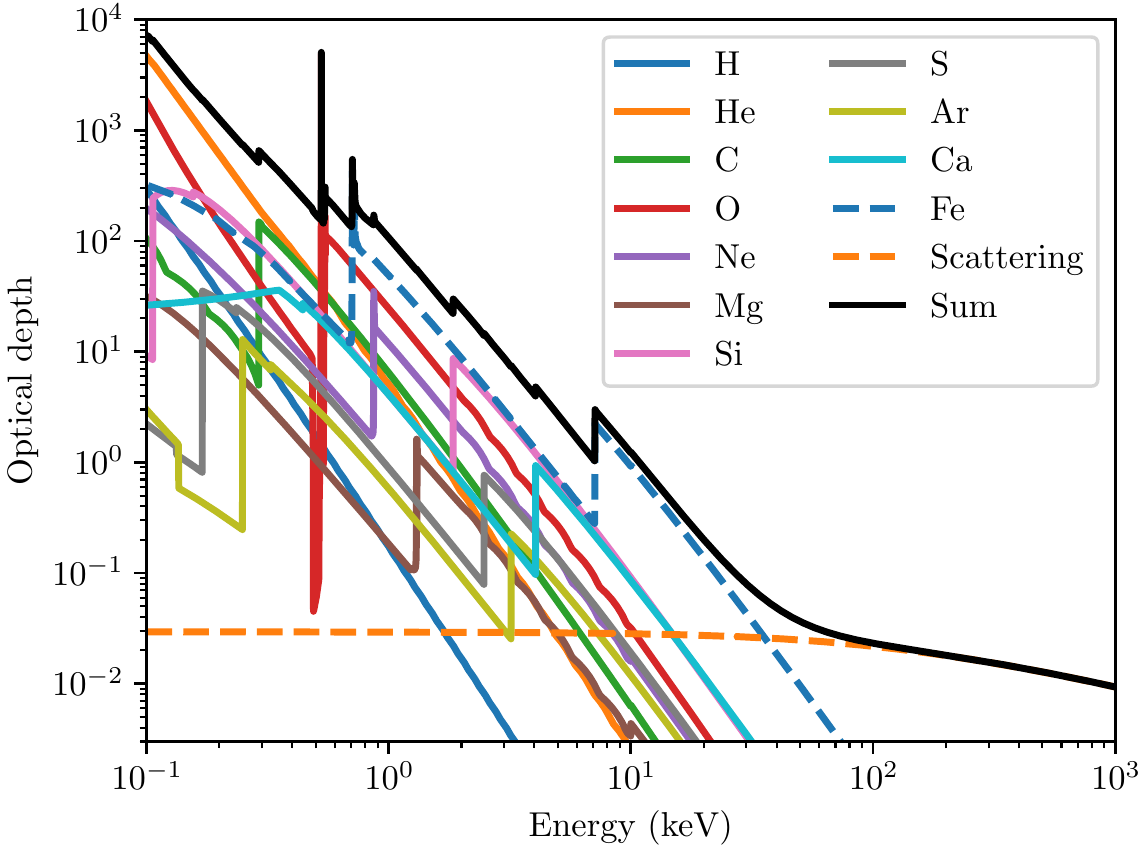}
  \caption{Direction-averaged optical depths for all models
    (left). The dashed lines are photoabsorption, the dotted lines are
    Compton scattering, and the solid lines are the sums. The blue B15
    line is partially covered by the green L15 line in the
    photoabsorption regime and the red W15 line in the scattering
    regime. The non-stripped models are very similar to each other and
    significantly different from the IIb model, which expels the
    ejecta at higher velocities because all but \henv{}\Msun{} of the
    hydrogen has been artificially removed. The individual
    contributions of all chemical elements to the optical depth for
    the B15 model (right). The individual contributions in all other
    models are qualitatively similar.\label{fig:avg_tau}}
\end{figure*}
Direction-averaged optical depths for all models and the individual
contributions for the B15 model at 10\,000 days are shown in
Figure~\ref{fig:avg_tau}. The optical depths are high and the ejecta
are opaque for soft X-rays for several decades after the explosion. A
typical optical depth for the non-stripped models is 30 at 2~keV.

The optical depth for the \iib{} model is approximately an order of
magnitude lower because the expansion velocity is higher, which is a
consequence of having removed all but \henv{}\Msun{} of the hydrogen
envelope. The decrease in optical depth because of the removed
hydrogen is subdominant because the contribution from hydrogen is
small even in the non-stripped models.

The general behavior is that He, C, O, Si, Ca, and Fe are all
dominating in different models within certain energy ranges. The
differences are determined by the relative abundances of the
metals. However, the relative abundances of the metals are uncertain
because of the small $\alpha{}$-networks used for the simulations
(Section~\ref{sec:unc_mod}).

\begin{figure*}
  \includegraphics[width=\columnwidth]{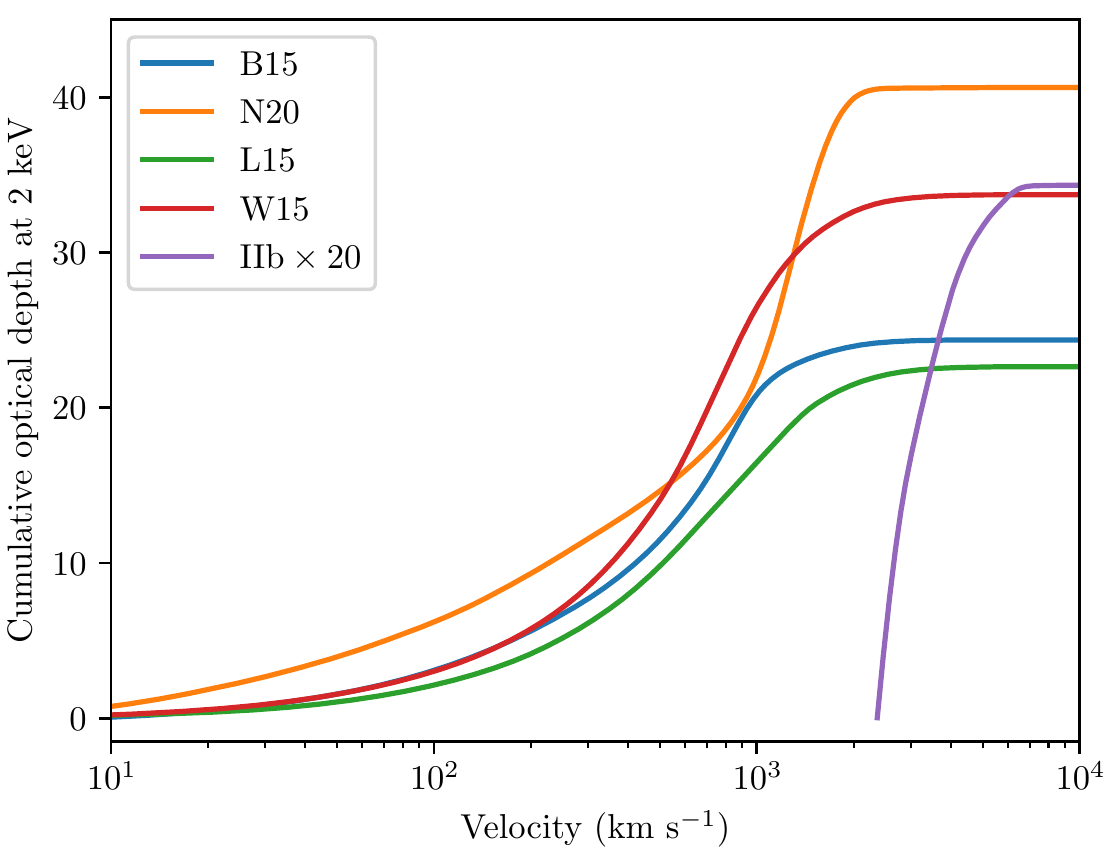}
  \hfill{}
  \includegraphics[width=\columnwidth]{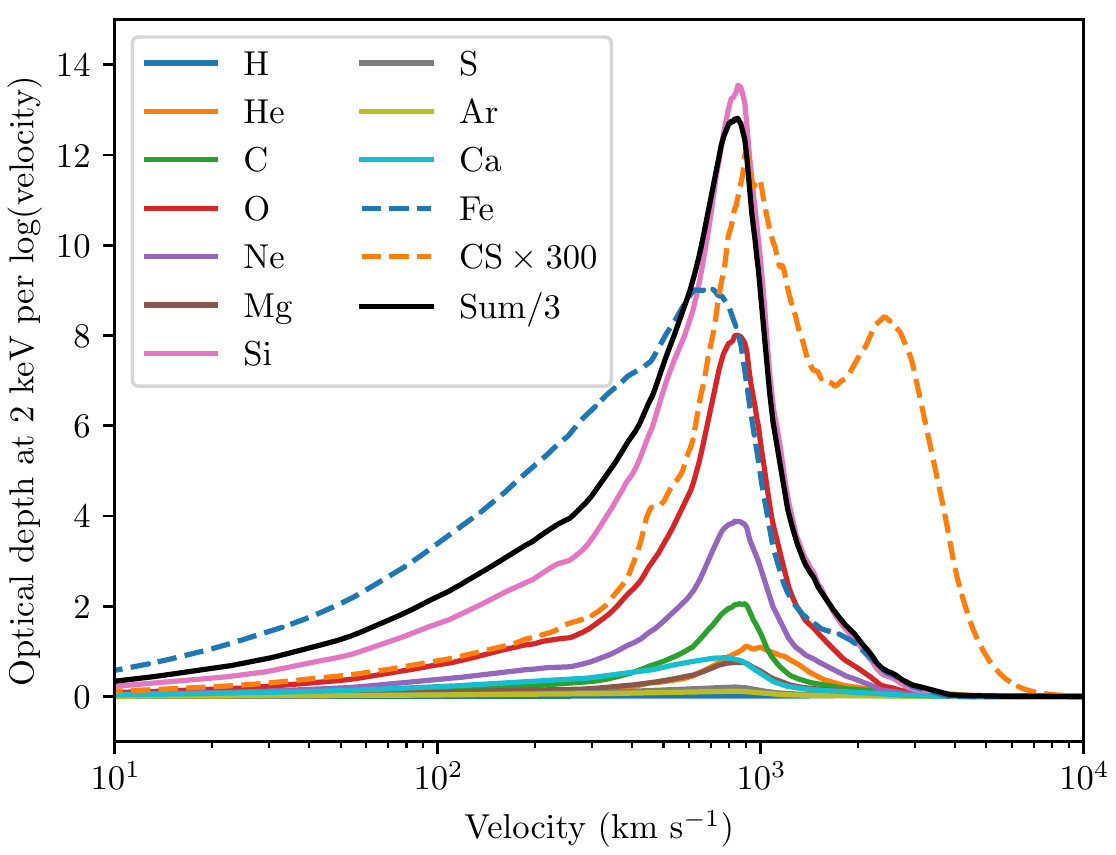}
  \caption{Cumulative direction-averaged optical depth at 10\,000 days
    at 2~keV as a function of velocity for all models (left). The
    optical depth of the \iib{} model has been scaled by a factor of
    20 for visual clarity. The effect of the cut off at 2400\kmps{}
    for the \iib{} model is expected to be small
    (Section~\ref{sec:methods}). The density gradient at the inner
    boundary of the \iib{} model is very steep because of a much
    weaker reverse shock, which is a consequence of the lack of a
    hydrogen envelope.  Contribution to the cumulative 2~keV optical
    depth per $\log(\mathrm{velocity})$ for the B15 model
    (right). Mathematically, this is
    $\mathrm{d}\,\tau(\text{2~keV})/\mathrm{d}\log(v)$, where $v$ is
    the velocity. The Compton scattering (CS) contribution and sum
    have been scaled for visual clarity. The bump at
    ${\sim}$2500\kmps{} in the scattering contribution is from
    hydrogen. The distribution is qualitatively similar for the other
    models.\label{fig:cum_tau}}
\end{figure*}
The cumulative direction-averaged optical depth at 2~keV as a function
of \replaced{radius}{velocity} is shown in Figure~\ref{fig:cum_tau}
(left) for all models. The dominant contribution to the optical depth
is from ejecta with velocities in the range 400--2000\kmps{} for the
non-stripped models. The velocity range is shifted to higher
velocities by a factor of ${\sim}$3 for the \iib{} model because of
the higher average ejecta velocity. We also analyze the contributions
to the cumulative distribution from individual elements
(Figure~\ref{fig:cum_tau}, right) and find that the general trend is
that the contributions from all metals are from approximately the same
\replaced{radii}{velocities}. There is a trend that contributions from
heavier elements are on average from \replaced{radii}{velocities}
lower than those of lighter elements. The effect of this is less than
a factor of a few in \replaced{radius}{velocity}, which is comparable
to the spread in \replaced{radii}{velocities} for individual elements.

\subsection{Asymmetries of the Ejecta}\label{sec:asy}
\begin{figure}
  \includegraphics[width=\columnwidth]{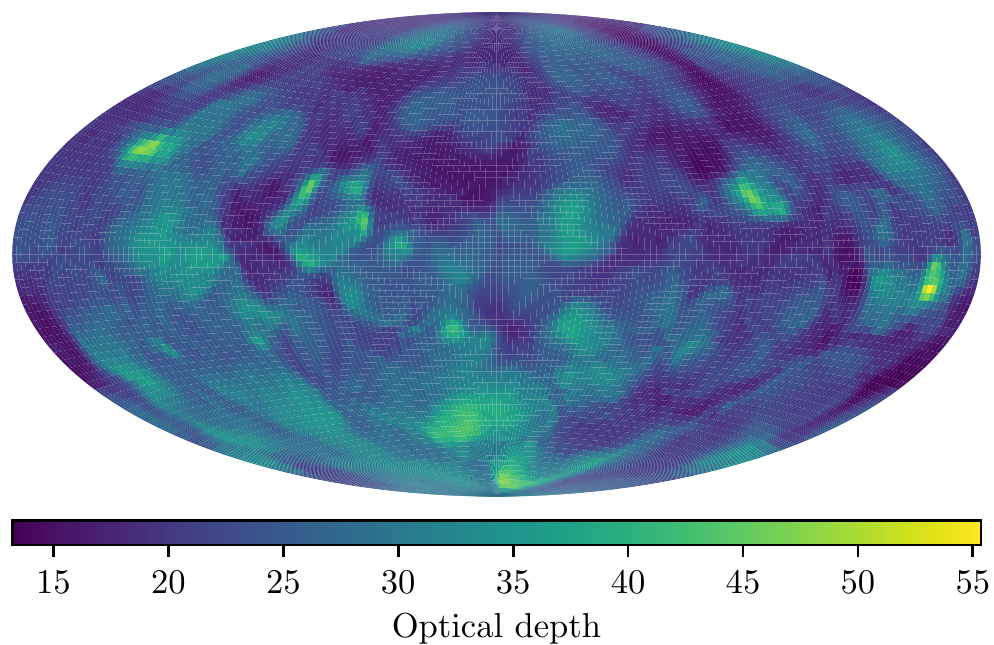}
  \includegraphics[width=\columnwidth]{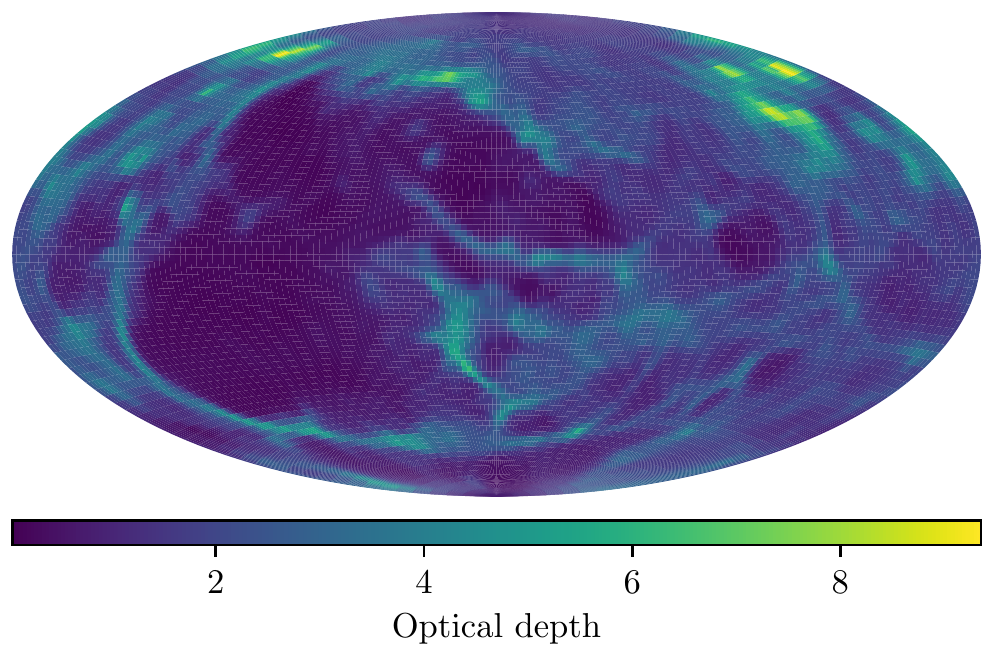}
  \caption{Equal-area Hammer projections of the optical depth at an
    energy of 2~keV of the B15 (top) and \iib{} (bottom) models at
    10\,000 days. The projection looks qualitatively similar at
    different energies for each of the models. The relative strength
    of the different structures that can be seen varies across
    absorption edges because of different
    compositions.\label{fig:tau_sky}}
\end{figure}
SN explosions are highly asymmetric and a consequence is that the
absorption depends on the line of sight
\added{(Section~\ref{sec:met_asy})}. The projections of the optical
depths for all directions of the B15 and \iib{} models are shown in
Figure~\ref{fig:tau_sky}. All models show pronounced asymmetric clumpy
structures or filaments. The variances in optical depths are caused by
a combination of varying amounts of ejecta expelled in different
directions and varying compositions of the ejecta.

We also visually inspected projections of column number densities of
individual elements (not shown). The most uniformly distributed
elements are the light elements H, He, and C whereas the heavy
elements Ca and Fe are most asymmetrically distributed. To a certain
extent, the directions of high abundances of heavy elements are
anti-correlated with light elements. We interpret this as rising
clumps of heavy elements that pierce through the outer shells of the
progenitor and leave holes or push the light elements into filaments.
It is often the case that the clumps consist of heavier elements
whereas the smoother variations and filaments are lighter elements
from the outer layers of the progenitor.

\begin{figure}
  \centering \includegraphics[width=\columnwidth]{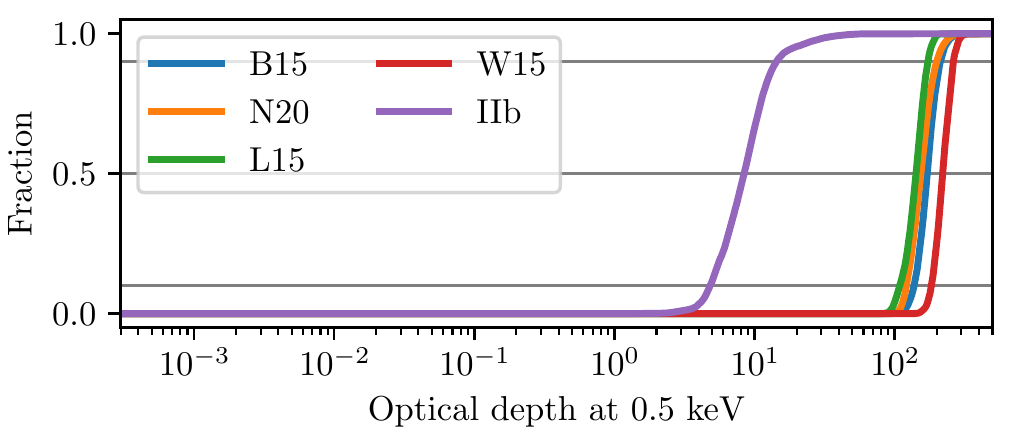}
  \centering \includegraphics[width=\columnwidth]{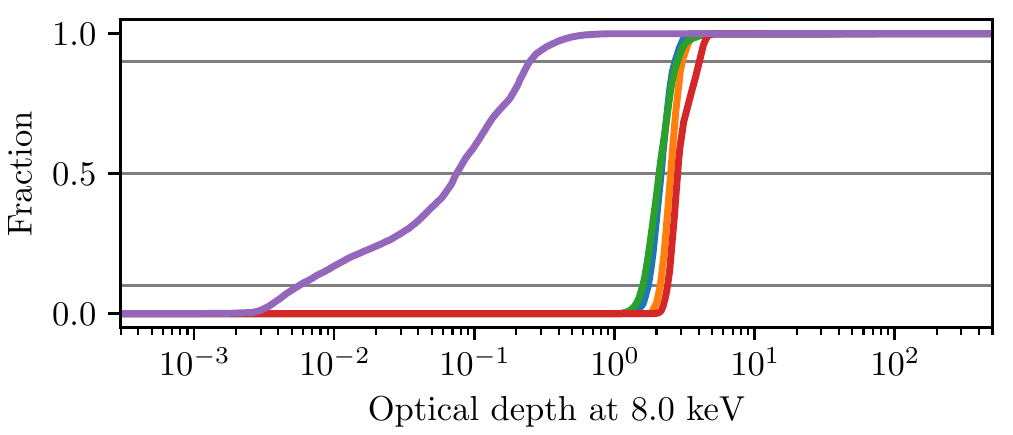}
  \centering \includegraphics[width=\columnwidth]{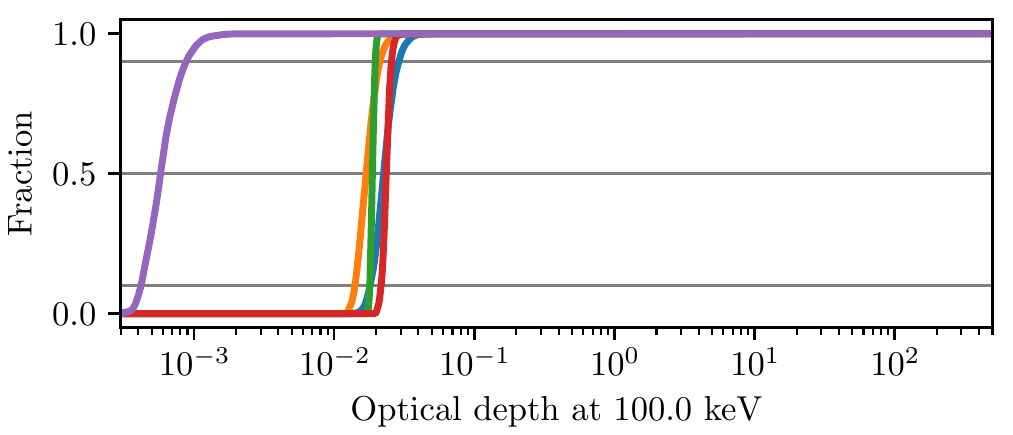}
  \caption{Cumulative distribution functions of the optical depth at
    10\,000 days as functions of direction. The horizontal gray lines
    show 90\,\%, 50\,\%, and 10\,\%. The non-stripped models are very
    similar, while the \iib{} model has lower optical depths. The
    iron-peak elements in the \iib{} model are also more
    asymmetrically distributed. This is clearly seen from the very
    wide CDF at 8~keV.\label{fig:tau_cdf}}
\end{figure}
Another representation of asymmetries is the cumulative distribution
function (CDF) of the optical depth. In the CDFs, the asymmetries are
correlated to the width of the distribution. The optical depth CDFs at
0.5, 8, and 100~keV are shown in Figure~\ref{fig:tau_cdf}. The
qualitative and quantitative behaviour of the non-stripped models are
similar. Optical depths of the \iib{} model are clearly lower but the
distribution of optical depths is also wider. The most notable feature
of the CDFs over all directions is the very wide distribution at 8~keV
for the \iib{} model, which is the result of highly asymmetric
distribution of iron-peak elements.

\begin{deluxetable*}{lccccccccccccccc}
  \tablecaption{Asymmetry Measures (1-$\sigma{}$ confidence intervals)\label{tab:tau_var}} \tablewidth{0pt}
  \tablehead{ \colhead{Model} &
    \colhead{$\tau_{0.5}$\tablenotemark{a}} &
    \colhead{$\tau_{1}$} & \colhead{$\tau_{2}$} &
    \colhead{$\tau_{4}$} & \colhead{$\tau_{8}$} & \colhead{$\tau_{100}$} &
    \colhead{$V_{0.5}$\tablenotemark{b}} &
    \colhead{$V_{1}$} & \colhead{$V_{2}$} &
    \colhead{$V_{4}$} & \colhead{$V_{8}$} & \colhead{$V_{100}$}}
  \startdata
  B15 & $171_{-26}^{+30}$ & $104_{-17}^{+24}$ & $24_{-5}^{ +6}$ & $ 4.0_{-0.8}^{+ 1.0}$ & $ 2.2_{-0.3}^{+ 0.4}$ &            $22.6_{-3.5}^{+4.6}\times{}10^{-3}$ & $ 1.4$ & $ 1.5$ & $ 1.6$ & $ 1.6$ & $ 1.4$ & $ 1.4$ \\
  N20 & $153_{-26}^{+35}$ & $175_{-41}^{+70}$ & $37_{-9}^{+16}$ & $ 6.0_{-1.3}^{+ 2.3}$ & $ 2.5_{-0.3}^{+ 0.4}$ &            $17.0_{-2.4}^{+3.2}\times{}10^{-3}$ & $ 1.5$ & $ 1.8$ & $ 1.9$ & $ 1.8$ & $ 1.3$ & $ 1.4$ \\
  L15 & $142_{-26}^{+26}$ & $112_{-28}^{+27}$ & $22_{-5}^{ +5}$ & $ 3.3_{-0.8}^{+ 0.8}$ & $ 2.1_{-0.4}^{+ 0.5}$ &            $18.8_{-0.6}^{+0.7}\times{}10^{-3}$ & $ 1.4$ & $ 1.6$ & $ 1.6$ & $ 1.6$ & $ 1.6$ & $ 1.1$ \\
  W15 & $221_{-30}^{+34}$ & $173_{-36}^{+39}$ & $33_{-6}^{ +7}$ & $ 4.9_{-0.9}^{+ 1.0}$ & $ 2.8_{-0.3}^{+ 1.0}$ &            $23.6_{-1.4}^{+1.6}\times{}10^{-3}$ & $ 1.3$ & $ 1.6$ & $ 1.5$ & $ 1.5$ & $ 1.5$ & $ 1.1$ \\
  IIb & $  8_{ -3}^{+ 4}$ & $  9_{ -6}^{ +6}$ & $ 2_{-1}^{ +1}$ & $ 0.2_{-0.2}^{+ 0.2}$ & $ 0.1_{-0.1}^{+ 0.1}$ & $\hphantom{0}0.6_{-0.1}^{+0.2}\times{}10^{-3}$ & $ 2.4$ & $ 4.9$ & $ 5.7$ & $ 6.0$ & $23.3$ & $ 1.8$ \\
  \enddata
  \tablenotetext{a}{$\tau_{E}$, optical depth at 10\,000~days at
    energy $E$ in keV.}
  
  \tablenotetext{b}{$V_{E}$, ratio of the 1\nobreakdash{-}$\sigma{}$
    upper to the 1\nobreakdash{-}$\sigma{}$ lower limit of
    $\tau_{E}$.}
  
  \tablecomments{The 1-$\sigma{}$ confidence intervals of the optical
    depths refers to the standard deviation of $\tau$ along different
    directions. This is effectively a measure of the asymmetry of the
    explosion model.}
\end{deluxetable*}
More quantitative measurements of the asymmetries are given by the
variations in optical depths at given energies for each model, which
are provided in Table~\ref{tab:tau_var}. The confidence intervals
represent the 1-$\sigma{}$ intervals of the optical depth as a
function of direction, which are at a level of approximately
20--40\,\%. We also include the ratio of the upper bound to the lower
bound as a relative measure of asymmetry variance. The optical depth
for a given model and energy typically spans a factor of 1.5--2. It is
also clear from the variance measures that the \iib{} model is more
asymmetric, in particular around 8~keV where the iron-peak elements
dominate the opacity.

\subsection{Ejecta Compositions}\label{sec:eje_com}
\begin{figure}
  \centering \includegraphics[width=\columnwidth]{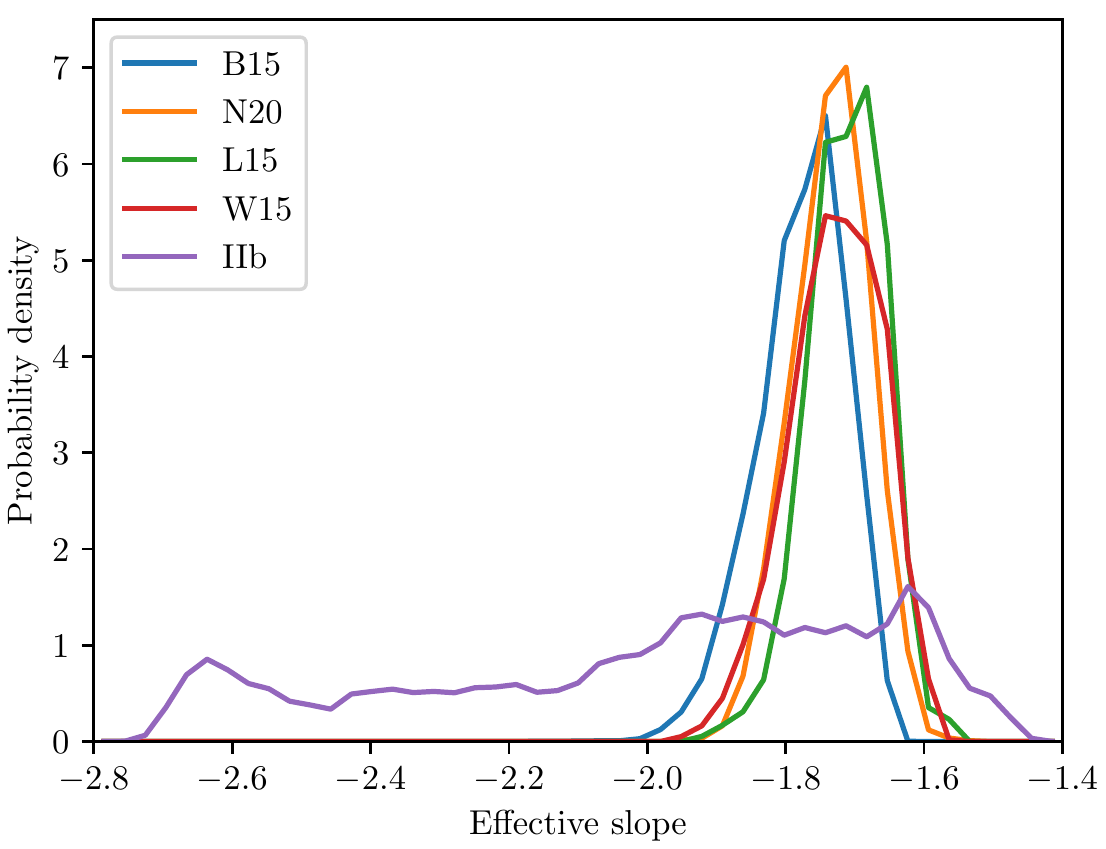}
  \caption{Probability distribution of the effective slope of the
    effective cross-section ($\alpha$) for all directions. The
    effective slope is defined by Equation~\eqref{eq:alpha}. Different
    compositions result in different effective cross-sections. Thus,
    the effective slope serves as a proxy for the
    composition.\label{fig:sha_dis}}
\end{figure}
The results up to this point have focused on the direction-averaged
optical depth and variance of the optical depth introduced by both
asymmetric distribution of material and composition. It is possible to
isolate the contribution to the variance of varying composition, which
affects the energy-dependence of the optical depth. We do not attempt
to quantify the distribution of abundances of all individual
elements. Instead, we define the effective slope $\alpha{}$ by
\begin{equation}
  \label{eq:alpha}
  \sigma_\mathrm{SN, eff}(\boldsymbol{\hat{n}}, E)\propto E^{\alpha},
\end{equation}
where $\sigma_\mathrm{SN, eff}$ is the effective cross-section of the
SN ejecta. The general trend is that lighter elements have steeper
energy dependencies and heavier elements have shallower
dependencies. Therefore, the effective slope can serve as a
representation of the variations in compositions as a function of
direction. The advantage of introducing $\alpha$ in place of the
composition is that the composition depends on eleven parameters (all
elements). The drawback is that it provides no insight into the
relative abundances of heavier elements, which determine the relative
strength of absorption edges. We choose to measure $\alpha$ as the
slope between 0.3 and 10~keV.

Figure~\ref{fig:sha_dis} shows the distribution of $\alpha$ for all
directions. The non-stripped models are similar and have distributions
that overlap almost completely, whereas the \iib{} model shows a wider
distribution of $\alpha{}$. This is a consequence of the asymmetric
distribution of iron-peak elements in the \iib{} model. For all
models, there is a significant spread in the distribution, which
indicates that the composition is varying. It is worth pointing out
that even though the slope is defined over the range 0.3--10~keV,
absorption is often only important in a limited interval where the
optical depth is around unity. This is because any lower opacity
leaves the spectrum relatively unaffected whereas higher depths are
likely to completely obscure the source.

\subsection{Compton Scattering}\label{sec:com_sca}
In contrast to photoabsorption, Compton scattering will not destroy
the photon, but will result in a down-scattering of the photon energy,
until photoabsorption dominates. This was discussed for the early
radioactive phase in \sna{} dominated by $^{56}$Ni and $^{57}$Ni
decay, as well as for pulsar input~\citep[e.g][and references
therein]{xu88, ebisuzaki88, grebenev88}. Here, we limit ourselves to
estimates of the optical depths and do not discuss the spectrum
resulting from the scattering further.

Figure~\ref{fig:avg_tau} shows the contribution of Compton scattering
to the optical depth. At 10\,000 days, the scattering depth above
50~keV is 1--$3\times 10^{-2}$ for the non-stripped models and
${\sim}5\times 10^{-4}$ for the \iib{} model. The dominating
contribution to the scattering depth is by hydrogen and helium. The
relative contributions can be seen from the electron column number
densities provided in Table~\ref{tab:abundances}. Assuming homologous
expansion, the scattering depth at 1~MeV is unity at ${\sim}$1000 days
for the non-stripped models and ${\sim}$150 days for the \iib{} model.

The total optical depths above 50~keV are dominated by Compton
scattering. The spread in scattering depths can be seen in
Figure~\ref{fig:tau_cdf} and Table~\ref{tab:tau_var}. The RSGs have
lower asymmetry variances in the scattering regime. The main reason
for this is that scattering is dominated by lighter elements, which
are more uniformly distributed in the more extended envelopes of the
RSGs.

\section{Discussion}\label{sec:discussion}
\subsection{Absorption Properties}
The absorption to the central regions of SNRs is high for X-rays for a
long time after the explosion. If the ejecta expand homologously, the
optical depth below 2~keV remains above unity for more than a century
for the non-stripped models and reaches unity at 10~keV at
approximately 30 years. The situation is very different for the \iib{}
model. Because the hydrogen envelope is stripped off, the SN explosion
expels the ejecta at a factor of 3 higher velocities, which results in
a factor of 9 lower optical depths. The contribution of hydrogen to
the optical depth constitutes less than ${\sim}10$\,\% of the total
photoabsorption depth above 0.3~keV for all models
(Figure~\ref{fig:avg_tau}).  The expansion velocity $v$ is also
connected to the SN explosion energy $E$ as $v \propto \sqrt{E/M}$,
where $M$ is the total ejecta mass. Therefore, it is reasonable to
assume that higher explosion energies or lower ejecta masses lead to
higher expansion velocities and lower optical depths.

Another important property of ejecta absorption is that the
metallicity is much higher than standard ISM abundances. This implies
that the hydrogen column density commonly used by X-ray astronomers is
a poor parameterization of the absorption. The effective cross-section
is much higher for high-metallicity gas than for standard ISM
compositions. The energy dependence of the effective cross-section is
significantly flatter for metal-rich than metal-poor gas, primarily
because high-metallicity gas has more pronounced absorption edges,
mostly K-shell edges, in the 0.3--10~keV energy range. The stronger
absorption edges are likely to be the most robust observational
signature of high-metallicity absorption, because the difference in
continuum absorption is likely degenerate with underlying model
components.

Compton scattering dominates the optical depth above 50~keV and the
energy dependence of the scattering depth is weak for energies lower
than 100~keV. The transition energy of 50~keV is higher than for ISM
compositions, where the transition occurs at ${\sim}$10~keV. The
reason for this is the higher metallicity of the SN ejecta, which
significantly increases the importance of photoabsorption and has a
much smaller effect on the scattering properties.

\subsection{Comparison with Previous Estimates}\label{sec:com}
In general, previous studies find SN ejecta absorption properties
comparable to our estimates. However, no previous work has explored 3D
SN explosion models, quantified the uncertainties introduced by SN
asymmetries, or compared different progenitors.

\citet{fransson87} modeled the absorption for X-rays in \sna{} using
the 15-M$_{\sun}$ 1D model 4A from \citet{woosley88}. They showed the
photoabsorption optical depth in their Figure~1 and found that the
optical depth reaches unity at 10~keV at 18~years. This can be
compared to our estimates for BSGs, which show that the optical depth
at 10~keV reaches unity at ${\sim}$30 years. However, they noted that
using the 25-M$_{\sun}$ model 8A~\citep{woosley88} resulted in optical
depths that are a factor of 2.2 higher, which better agrees with our
results. \citet{fransson87} also concluded that Compton scattering
starts dominating over photoabsorption in the 30--100~keV range,
depending on the composition of the gas.

\citet{serafimovich04} used the 15-M$_{\sun}$ 1D model 14E1 of
\citet{blinnikov00} to estimate the ejecta absorption in the
1000-year-old pulsar PSR B0540-69.3 in the Large Magellanic
Cloud. They reported that the effective cross-section for SN ejecta is
40 times higher than for standard ISM abundances (see
Appendix~\ref{app:com}) and that the effect of ejecta absorption is
negligible because of the age of PSR B0540-69.3, which agrees with our
results.

\citet{orlando15} performed simulations describing \sna{} and found an
ejecta hydrogen column density of ${\sim}3\times 10^{22}$~cm$^{-2}$ at
30 years. This number is close to our values of the hydrogen column
number densities (Table~\ref{tab:abundances}). However, they do not
report the chemical composition of the ejecta, which is critical for a
complete absorption estimate.

\citet{esposito18} used the model lm18a7Ad of \citet[][provided to
them by S.~E. Woosley]{dessart10}. It is a 1D model with a total
ejecta mass of 15.6\Msun{}. \citet{esposito18} used the abundances
from the model, assumed homogeneous distribution, and estimated the
density based on approximations. They reported a hydrogen column
density of $7.6\times 10^{21}$~cm$^{-2}$ for the ejecta, which should
result in absorption properties comparable to our estimates for the
abundances provided in their Table~2.

\subsection{Sources of Uncertainty}\label{sec:errors}
There are several sources of uncertainty that we have neglected. The
presented uncertainties on the optical depths in
Table~\ref{tab:tau_var} only represent the variance introduced by the
asymmetries for a given model. One of the purposes of providing column
densities for comparable models is to get a handle on the
uncertainties associated with the progenitor properties, stellar
evolution, and explosion simulations. The sensitivity of the results
to progenitor models and early explosion simulations can be estimated
by comparing the four non-stripped models. They were not tuned to
represent the same progenitor, only B15 and N20 were designed to match
the progenitor of \sna{} in broad terms~\citep[see][]{utrobin15}. Our
results show that the general absorption properties of all
non-stripped models are relatively similar. This is most clearly seen
from the CDFs in Figure~\ref{fig:tau_cdf} and the values in
Table~\ref{tab:tau_var}. The differences between the
direction-averaged properties of the models are comparable to the
asymmetry uncertainties within each model. This indicates that the
effects of progenitor properties and asymmetry uncertainties are
comparable for a given SN type.

Observations of nearby SNRs find dust masses of ${\sim}0.1$\Msun{} or
larger in \cas{}~\citep{rho08, barlow10}, the Crab
nebula~\citep{gomez12, owen15}, and \sna{}~\citep{matsuura11,
  matsuura15, indebetouw14, dwek15}. Molecules composed of metals have
also been detected in \cas{}~\citep{rho12}, the Crab
nebula~\citep{barlow13, bentley18}, and \sna{}~\citep{kamenetzky13,
  matsuura17, abellan17}. However, the formation of dust and molecules
from the ejecta material only has a small effect on the X-ray
absorption properties~\citep{morrison83, draine03}. Photoabsorption in
soft X-rays is dominated by inner-shell absorption, which is
approximately the same for dust and molecules as their constituents
and Compton scattering only depends on the electron column density.

Additionally, we do not consider the effects of the compact object
kick velocity. A full treatment would require investigation of
absorption to all positions that the compact object can have been
kicked to, but the 3D kick velocity is rarely known. Some insight into
the possible effects of kick velocities on absorption is provided by
Figure~\ref{fig:cum_tau}, which shows at what radii the gas
contributes to the optical depth. For kick velocities of
${\sim}500$\kmps{}, the maximum effect on the optical depth is
${\sim}50$\,\% if the kick happens to be aligned with our
line-of-sight. For reference, we note that pulsars are inferred from
observations to have 3D kick velocities of
${\sim}400\pm{}250$\kmps{}~\citep{hobbs05, faucher-giguere06} with
some extreme cases up to 1600\kmps{}~\citep{cordes93, chatterjee02,
  chatterjee04, hobbs05}.

\subsubsection{Progenitor and Explosion Models}\label{sec:unc_mod}
All results rely on the SN progenitors and explosion models. These are
not fully self-consistent and it is difficult to exactly quantify the
uncertainties introduced by them. We find it unlikely that the model
uncertainties have a major, qualitative impact on our conclusions.
What follows is a brief description of some of the contributing
factors.

The explosion models that we use do not include a longer-lasting phase
of simultaneous accretion and mass outflow around the NS after the
onset of the explosion as suggested to exist by present
self-consistent models~\citep{muller17}. Having such a phase of
accretion and outflow instead of the spherical ``wind'' that boosts
the explosion in the simulations, is likely to lead to more extreme
asymmetries in the innermost ${\sim}$0.1\Msun{} of the ejecta.
Therefore, the inner half of the iron-group matter is likely to be
more asymmetrically distributed. Additionally, the 3D simulations
contain no information about mixing below the grid scale, but this is
not important for the X-ray absorption properties because the
absorption only rely on the total column densities.

Overall, the explosion dynamics is roughly compatible with our
present-day knowledge of how neutrino-driven explosions
work. Moreover, the asymmetries we find in the simulations are able to
produce the mixing needed to explain the light curve of
\sna{}~\citep{utrobin15}, the morphology of
\cas{}~\citep{wongwathanarat17}, or, roughly, the asymmetries of SiO
and CO molecules observed in \sna{}~\citep{abellan17}. However, an
important caveat for the \sna{} models is that they are based on
single-star progenitors. The progenitor of \sna{} was most likely the
result of a merger, which is primarily supported by the triple-ring
structure surrounding the SN~\citep{blondin93, morris07, menon17,
  urushibata18}. The mixing properties of the merger models are
different and the mass of the helium and oxygen cores are also much
lower than for the single-star progenitors. However, the set of
progenitors considered in our study covers a fairly wide range of
pre-collapse profiles, and one case (B15) has a core structure with
similarities to the core properties of some of the binary models of
\citet{menon17}.

The relative abundances of the different metals are also uncertain
because of the small $\alpha{}$\nobreakdash{-}nuclei
networks~\citep[see Section~3.1 of][]{wongwathanarat17}. This
uncertainty only applies to the relative abundances of metals and
leaves the total mass and spatial distribution unaffected.

Model B15, despite its ability to yield efficient mixing and to
explain the \sna{} light curve fairly well, is far from being a
perfect case for \sna{}. In particular, the NS mass is too low. A
baryonic mass of 1.25\Msun{} corresponds to a gravitational mass of
only ${\sim}$1.15\Msun{}. It is extremely unlikely that \sna{}
produced a NS with such a record-low mass~\citep[cf.][]{ozel16}. This
problem is most likely due to the progenitor structure for B15 rather
than the explosion modeling.

\subsection{Implications for Observations}\label{sec:imp}
The high X-ray absorption in young SNRs affects interpretations of
observations. We have focused on \cas{} and \sna{}, but the absorption
estimates also apply to other extragalactic SNRs. Processes that
produce X-rays from young compact objects are thermal surface
emission, fallback accretion, and pulsar wind activity. Ejecta
absorption obscures any X-ray emission and is generally applicable,
but it is possible that extended sources, such as powerful pulsar wind
nebulae, produce a significant amount of radiation from outer regions
where the absorption is expected to be lower. Another remark on pulsar
wind nebulae is that they produce synchrotron emission that extends to
high X-ray energies~\citep[for a high-energy compilation of the pulsed
emission, see Figure~28 of][]{kuiper15} where Compton scattering is
dominating. Given the low scattering depths, the obscuration due to
scattering is most likely only important during the first few years.

Another source of high-energy radiation is from radioactive elements
synthesized in SN explosions that emit lines in the Compton absorption
regime when decaying. This was seen in early observations of
\sna{}~\citep{sunyaev87, dotani87, matz88}. The first sign of
radioactive line emission at intermediate optical depths is the
emergence of down-scattered X-ray emission, which is then followed by
the direct line emission when the ejecta has expanded further. The
time and escaping fluxes of the radioactive emission convey
information about the level of mixing of the radioactive elements and
could be used to test the accuracy of the SN explosion simulations.

Ejecta absorption can potentially also explain the lack of detection
of K-shell emission lines resulting from electron capture in
radioactive elements produced by the SN~\citep{leising01, leising06,
  theiling06}. Radioactive K-shell lines have been observed in
G1.9+0.3, which is the remnant of a thermonuclear
SN~\citep{borkowski10}. Because of the lower mass and higher expansion
velocity of Type~Ia SNe, the X-ray absorption should be considerably
smaller for a given age.

We focus on absorption toward the center of SNRs. The absorption along
lines of sight that pierce outer parts of the ejecta have different
properties. This is mostly relevant for spatially resolved studies of
SNRs. Figure~\ref{fig:cum_tau} provides some insight into the radial
distribution of absorbing gas. We do not attempt to investigate all
possible sightlines. The difference introduced by shifting line of
sight is not simply a matter of geometrical scaling because the
metallicity is higher for paths into the center and lower toward the
outer boundary of the ejecta. An effect of this in \sna{} can be seen
in Figures~22 and~23 of \citet{fransson13}. The penetration of the
X-rays from the ring interaction is considerably deeper in the
directions of the outer hydrogen envelope compared to the radial
penetration in the core direction.

In \sna{}, the absorption is high and significantly affects
observations of the compact object. The analysis is presented in
detail in a separate work~\citep{alp18}. Here we note that the
differences in optical depth for the different BSG progenitor models
B15 and N20 are less than a factor of 2 and the RSGs also show
comparable properties (Section~\ref{sec:results}). A difference of a
factor of 2 in the optical depth approximately corresponds to a factor
of two in the allowed luminosities. This uncertainty is comparable to
the other uncertainties discussed in Section~\ref{sec:errors}.

The effect of absorption is expected to be relatively small for the
Galactic SNRs. We find a negligible effect of absorption on the
interpretation of \textit{Chandra} observations of the CCO in \cas{}
(Appendix~\ref{app:cas}). However, observations of SNRs with higher
number of total counts or higher energy resolution should be able to
disentangle small effects of ejecta absorption. Our estimates also
show that the low-energy X-ray emission from a compact object in a
future Galactic SN will be heavily obscured for many decades. It is
unlikely that even future telescopes will be able to directly observe
emission from newly created compact objects at energies less than
${\sim}5$~keV during the first few decades unless the compact object
is very luminous, such as the Crab~\citep{buhler14}.

\section{Summary \& Conclusions}\label{sec:conclusions}
Accurate models of absorption are critical for analyses of X-ray
observations of SNRs. We use 3D simulations of neutrino-driven SN
explosions~\citep{wongwathanarat13, wongwathanarat15} to estimate the
column densities of the most abundant elements along the line of sight
to the center of the ejecta.  The column densities are used to compute
the optical depth for X-rays due to photoabsorption and Compton
scattering. This provides both the amount and composition of the
obscuring gas. We use our absorption models to place new X-ray limits
on the compact object in \sna{} in an accompanying
paper~\citep{alp18}, and re-analyze the X-ray spectrum of the CCO in
\cas{} (Appendix~\ref{app:cas}). Our main conclusions are the
following:

\begin{itemize}
\item The optical depth for X-rays is high for a long time after the
  SN explosion. For the models with a hydrogen envelope, the optical
  depth between 0.1 and 50~keV is approximately given by
  $100\,t_4^{-2}\,E^{-2}$, where $t_4$ is time since the explosion in
  units of 10\,000 days (${\sim}$27 years) and $E$ the energy in units
  of keV.
\item The optical depth above ${\sim}$50~keV is dominated by Compton
  scattering and is 1--$3\times{}10^{-2}$ at 10\,000 days for models
  with a hydrogen envelope. Scaling backward in time, the scattering
  depth to the center is unity at around 1000 days after the
  explosion.
\item The optical depth is approximately an order of magnitude lower
  for the IIb model, which has lost all but \henv{}\Msun{} of its
  hydrogen envelope. This model explodes as a Type~IIb SN and shows
  similarity to \cas{}. The optical depth is lower because $\tau
  \propto 1/(vt)^2 \propto M/Et^2$.
\item The expected level of absorption in \cas{} based on estimates
  using the IIb model results in a decrease in the inferred CCO
  surface temperature of less than ${\sim}1$\,\% at all epochs, which
  is lower than the statistical uncertainty. We also find that the
  ejecta absorption component would be degenerate with the ISM
  absorption, if ejecta absorption was significant. This is a result
  of the limited statics of the observation, instrumental energy
  resolution, and degeneracy with parameters of other model
  components.
\item For the same hydrogen column number density, the metallicity of
  the SN ejecta is ${\sim}$2 orders of magnitude higher than for the
  ISM. This implies that the hydrogen column number density is a poor
  measure of absorption because the effective cross-section is much
  higher. The high metallicity results in a flatter energy dependence
  of the cross-section and that the absorption profile has stronger
  metal edges.
\end{itemize}

\acknowledgments{\added{We thank the anonymous referee for the helpful
    comments.} This research was funded by the Knut \& Alice
  Wallenberg Foundation and the Swedish Research Council. At Garching,
  this research was supported by the Deutsche Forschungsgemeinschaft
  through Excellence Cluster Universe (EXC 153;
  http://www.universe-cluster.de/) and Sonderforschungsbereich SFB
  1258 "Neutrinos and Dark Matter in Astro- and Particle Physics", and
  by the European Research Council through grant ERC-AdG
  No. 341157-COCO2CASA. The computations of the supernova models were
  carried out on Hydra of the Max Planck Computing and Data Facility
  (MPCDF) Garching and on the Cluster of the Computational Center for
  Particle and Astrophysics (C2PAP) Garching. The scientific results
  reported in this article are based in part on data obtained from the
  \textit{Chandra} Data Archive. This research has made use of
  software provided by the \textit{Chandra} X-ray Center (CXC) in the
  application package CIAO. This research has made use of NASA's
  Astrophysics Data System.}

\vspace{5mm}
\facilities{\textit{CXO}(ACIS)}
\software{astropy~\citep{astropy13},
  CIAO/CALDB~\citep{fruscione06},
  \deleted{IRAF/PyRAF,}
  \texttt{matplotlib}~\citep{hunter07},
  MARX~\citep{davis12}, 
  \texttt{numpy}~\citep{jones01, van_der_walt11},
  \added{\textsc{Prometheus}~\citep{fryxell91, muller91},
    \textsc{Prometheus-HOTB},
  \textsc{Prometheus-Vertex}~\citep{rampp02},}
  \texttt{scipy}~\citep{jones01},
  XSPEC~\citep{arnaud96}
}

\appendix
\section{\cas{}}\label{app:cas}
\subsection{Observations and Data Reduction}\label{sec:cas_obs}
\begin{deluxetable}{cccccccccc}
  \tablecaption{\textit{Chandra} \cas{} CCO Observations\label{tab:cas_obs}}
  \tablewidth{0pt}
  \tablehead{\colhead{Obs. ID} & \colhead{Start date} & \colhead{Exposure} &
    \colhead{Source counts\tablenotemark{a}} & \colhead{Source fraction\tablenotemark{b}} \\
    \colhead{} & \colhead{(YYYY-mm-dd)} & \colhead{(ks)} & \colhead{} & \colhead{}} \startdata
  \hphantom{0}6690 & 2006-10-19 & 61.7 & 6261 & 0.958 \\
             13783 & 2012-05-05 & 63.4 & 5571 & 0.953 \\
             16946 & 2015-04-27 & 68.1 & 5320 & 0.954 \\
  \enddata
  \tablenotetext{a}{Defined as the number of counts extracted from the
    source region.}
  \tablenotetext{b}{Defined as the fraction of the source counts that
    are signal, whereas the rest is background.}
\end{deluxetable}
We analyze the three archival \textit{Chandra} observations that were
performed using an instrumental setup chosen to minimize spectral
distortions of the CCO. Details of these observations are presented in
Table~\ref{tab:cas_obs}. We reduce the data using CIAO 4.9 and CALDB
4.7.7~\citep{fruscione06} and follow standard data reduction
guidelines. No strong background flares are detected and the data are
filtered by removing all intervals that deviate from the mean by more
than 4$\sigma{}$. All source spectra are extracted from circular
regions with radii of 2 pixels (492~mas~pixel$^{-1}$). All background
spectra are extracted from annuli centered on the source position. The
inner radii of the annuli are 3 pixels and the outer 5 pixels. The
spectra are binned such that the signal-to-noise level is at least 10
in each bin. Spectral fitting is performed using XSPEC version
12.9.1p~\citep{arnaud96}.

\subsection{Results}
Our aim is to investigate if ejecta absorption has a significant
effect on the interpretation of the observed spectra of the CCO in
\cas{}. To do this, we choose a model consisting of thermal emission
from a NS surface with a carbon atmosphere, which is modeled using the
XSPEC model \texttt{carbatm}~\citep{suleimanov14}. We use the model
\texttt{tbabs}~\citep{wilms00} with \texttt{wilm} abundances from
\citet{wilms00} for the ISM absorption. The ejecta absorption is
modeled using the XSPEC model \texttt{tbvarabs}~\citep{wilms00}, which
allows for setting the individual abundances. We take the column
densities of the \iib{} model (Table~\ref{tab:abundances}) because it
shows similarity to \cas{}~\citep{wongwathanarat17}. All column
densities are scaled to 340 years, assuming homologous expansion and
ignoring the SNR age difference of the observations. To study an
extreme scenario, we scale the average column densities by a factor of
4.1. This value is the ratio of the 99.7$^\mathrm{th}$ percentile of
the optical depth at 2~keV to the direction-averaged optical depth at
2~keV. This can be interpreted as a 3-$\sigma$ upper limit on the
ejecta absorption for the \iib{} model. The result of scaling to an
age of 340 years and taking the 3-$\sigma{}$ upper limit is that the
optical depth at 1~keV is ${\sim}$0.2, implying that the total effect
of ejecta absorption is expected to be small.

\begin{deluxetable}{ccccccc}
  \tablecaption{\cas{} CCO Fitting Results (1-$\sigma{}$ confidence intervals)\label{tab:cas_fit}}
  \tablewidth{0pt}
  \tablehead{\colhead{Obs.} & \colhead{$N_\mathrm{ISM}(\mathrm{H})$\tablenotemark{a}} &
    \colhead{$\tau(1\text{~keV})$} & \colhead{$T$\tablenotemark{c}} & \colhead{Norm.\tablenotemark{a}} & \colhead{$\chi^2/\text{d.o.f.}=\chi^2_\mathrm{red}$} \\
    \colhead{(YYYY)} & \colhead{($10^{22}$~cm$^{-2}$)} & \colhead{} & \colhead{(MK)} & \colhead{} & \colhead{}} \startdata
  \multicolumn{6}{c}{{{No ejecta absorption}}} \\\hline
  2006 &                        &           & $2.07^{+0.05}_{-0.05}$ &                     &                \\
  2012 & $2.15^{+0.05}_{-0.05}$ & \nodata{} & $2.05^{+0.05}_{-0.05}$ & $7.3^{+1.2}_{-1.0}$ & $118/134=0.88$ \\
  2015 &                        &           & $2.03^{+0.05}_{-0.05}$ &                     &                \\\hline
  \multicolumn{6}{c}{{{3-$\sigma{}$ ejecta absorption}}}\\\hline
  2006 &                        &     & $2.06^{+0.05}_{-0.04}$ &                      &                \\
  2012 & $2.01^{+0.05}_{-0.05}$ & 0.2 & $2.04^{+0.05}_{-0.04}$ &  $7.6^{+1.1}_{-1.1}$ & $119/134=0.89$ \\
  2015 &                        &     & $2.02^{+0.05}_{-0.04}$ &                      &                \\\hline
  \multicolumn{6}{c}{{{$10\times3$-$\sigma{}$ ejecta absorption}}}\\\hline
  2006 &                        &     & $1.99^{+0.05}_{-0.05}$ &                      &                \\
  2012 & $0.77^{+0.05}_{-0.05}$ & 2.3 & $1.97^{+0.05}_{-0.04}$ &  $9.7^{+1.6}_{-1.3}$ & $123/134=0.92$ \\
  2015 &                        &     & $1.95^{+0.04}_{-0.04}$ &                      &                \\
  \enddata
  \tablenotetext{a}{Free but tied across observations.}
  \tablenotetext{b}{Frozen to IIb model prediction.}
  \tablenotetext{c}{Local (unredshifted) temperature.}
\end{deluxetable}
We perform the fit of the model simultaneously to the three
observations. The results of all fits are presented in
Table~\ref{tab:cas_fit}. First, we start by ignoring any ejecta
absorption. The NS mass is frozen to $1.647$\Msun{} and the radius is
frozen to $10.33$~km, to allow for comparisons with previous
works~\citep{heinke10, posselt13}. The normalization of the NS carbon
atmosphere model is a ratio of the fraction of the surface that is
emitting ($A$) over the distance squared in units of 10~kpc. We leave
the normalization free to allow for any normalization errors, but tied
it across all observations because we do not expect the emitting
fraction or distance to change. For a distance of 3.4~kpc, the
normalization is 8.65 for $A=1$. The ISM hydrogen column density
$N_\mathrm{ISM}(\mathrm{H})$ is left free because the absorption
toward \cas{} is uncertain~\citep{keohane96, salas18}, but it is tied
across observations because it is not expected to vary significantly
on these timescales. Additionally, we find no statistically
significant improvement by leaving $N_\mathrm{ISM}(\mathrm{H})$
untied. The local (unredshifted) blackbody temperature ($T$) is
left free.

Next, we include the extreme-case 3-$\sigma{}$ ejecta absorption. The
best-fit parameters are shown in the second segment of
Table~\ref{tab:cas_fit}. The additional ejecta absorption component
has no significant impact and the fit remains essentially unchanged.
This is expected because of the low optical depth of ${\sim}$0.2 at
1~keV, which implies that the difference should be very small.

To explore a case of significant absorption, we add another factor of
10 to the 3-$\sigma{}$ absorption. The factor of 10 is arbitrarily
chosen but the scenario could represent a case where the IIb model
overestimates the bulk expansion velocity or if filamentary structures
of higher density have formed along the line of sight. The best-fit
values for this case are also provided in Table~\ref{tab:cas_fit}. The
fit is statistically slightly worse but still acceptable. The
additional ejecta absorption is nearly completely degenerate with the
other components, particularly the ISM absorption, but also the
temperature and normalization to a certain extent.

The conclusion is that the expected level of ejecta absorption based
on our models has an insignificant effect on the estimated temperature
of the CCO. The difference in inferred temperature is less than
${\sim}1$\,\% between the fit without and with 3-$\sigma{}$ ejecta
absorption, which is lower than the statistical
uncertainty. Additionally, the effect is a similar level of decrease
at all epochs, which does not affect investigations of the cooling of
the CCO.

\section{Comparison with ISM Absorption}\label{app:com}
\begin{figure}
  \centering \includegraphics[width=0.48\columnwidth]{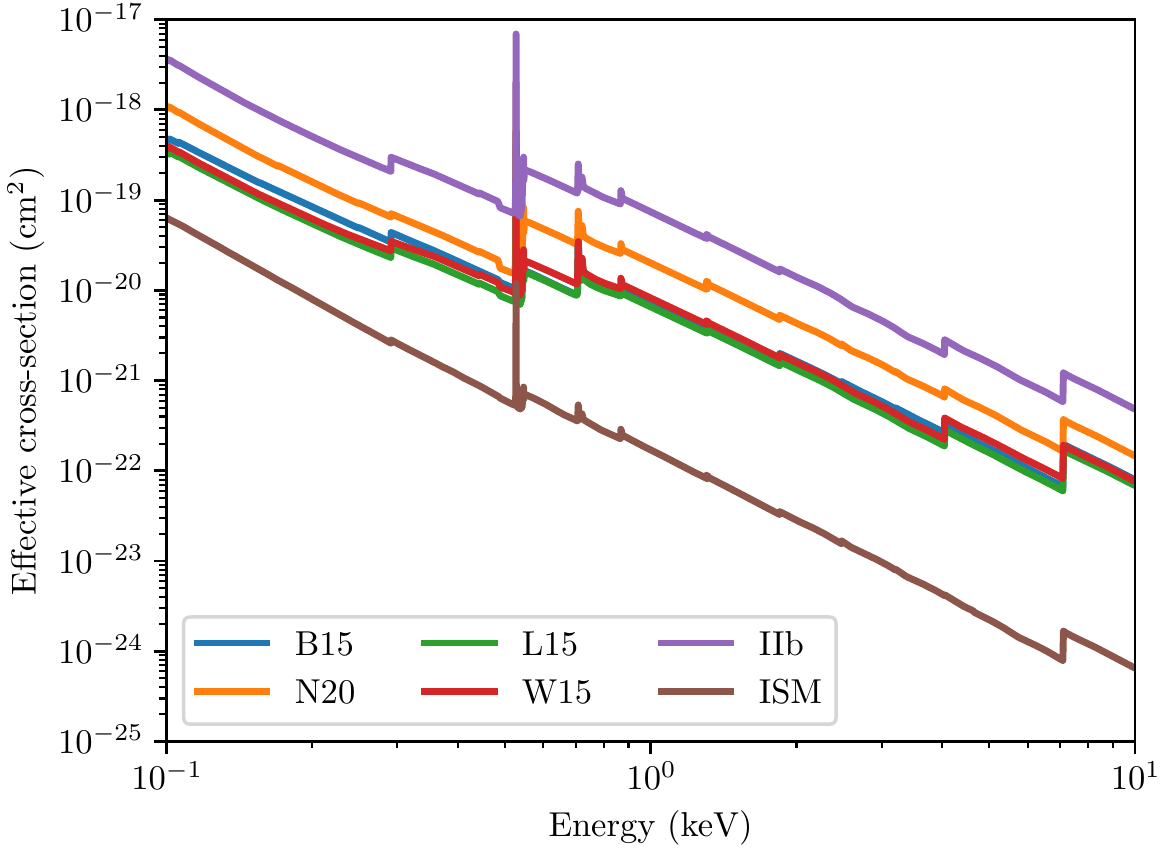}
  \hfill{}
  \centering \includegraphics[width=0.48\columnwidth]{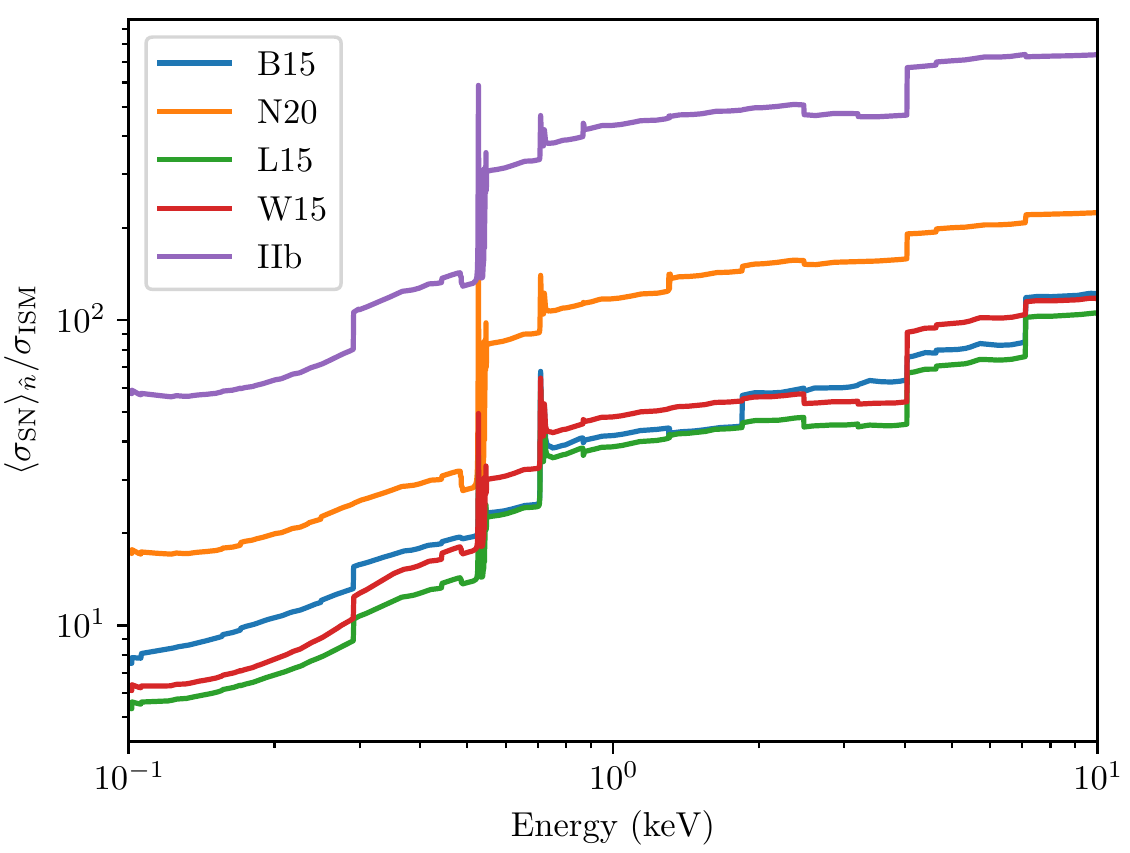}
  \caption{Comparison of the effective cross-sections of the SN ejecta
    for the five models with the ISM cross-section. The left panel
    shows the cross-sections as functions of energy and the right
    panel shows the ratio of the ejecta cross-sections to the ISM
    cross-section. We note that the SN effective cross-sections are
    much larger, show shallower energy dependencies, and have much
    stronger discontinuous absorption features.\label{fig:ism_com}}
\end{figure}
A comparison of the SN abundances with the ISM
abundances~\citep{wilms00} is provided in
Table~\ref{tab:abundances}. It is common in X-ray astronomy to
quantify absorption by the hydrogen column number density and to
assume standard abundances of all other elements with respect to
hydrogen. This assumption breaks down for ejecta absorption because
the metallicity of the SN ejecta is much higher than that of the
ISM. The result is that the effective cross-section of the SN ejecta
is much higher, falls of slower as a function of energy, and that
metal absorption edges are stronger. The effective cross-sections for
the five models and the ISM, as well as the ratio of the SN ejecta to
ISM cross-sections are shown in Figure~\ref{fig:ism_com}.

\bibliography{references}

\begin{thebibliography}{}
\expandafter\ifx\csname natexlab\endcsname\relax\def\natexlab#1{#1}\fi
\providecommand{\url}[1]{\href{#1}{#1}}
\providecommand{\dodoi}[1]{doi:~\href{http://doi.org/#1}{\nolinkurl{#1}}}
\providecommand{\doeprint}[1]{\href{http://ascl.net/#1}{\nolinkurl{http://ascl.net/#1}}}
\providecommand{\doarXiv}[1]{\href{https://arxiv.org/abs/#1}{\nolinkurl{https://arxiv.org/abs/#1}}}

\bibitem[{{Abell{\'a}n} {et~al.}(2017){Abell{\'a}n}, {Indebetouw}, {Marcaide},
  {Gabler}, {Fransson}, {Spyromilio}, {Burrows}, {Chevalier}, {Cigan},
  {Gaensler}, {Gomez}, {Janka}, {Kirshner}, {Larsson}, {Lundqvist}, {Matsuura},
  {McCray}, {Ng}, {Park}, {Roche}, {Staveley-Smith}, {van Loon}, {Wheeler}, \&
  {Woosley}}]{abellan17}
{Abell{\'a}n}, F.~J., {Indebetouw}, R., {Marcaide}, J.~M., {et~al.} 2017,
  \apjl, 842, L24, \dodoi{10.3847/2041-8213/aa784c}

\bibitem[{{Alp} {et~al.}(2018){Alp}, {Larsson}, {Fransson}, {Indebetouw},
  {Jerkstrand}, {Ahola}, {Burrows}, {Challis}, {Cigan}, {Cikota}, {Kirshner},
  {van Loon}, {Mattila}, {Ng}, {Park}, {Spyromilio}, {Woosley}, {Baes},
  {Bouchet}, {Chevalier}, {Frank}, {Gaensler}, {Gomez}, {Janka}, {Leibundgut},
  {Lundqvist}, {Marcaide}, {Matsuura}, {Sollerman}, {Sonneborn},
  {Staveley-Smith}, {Zanardo}, {Gabler}, {Taddia}, \& {Wheeler}}]{alp18}
{Alp}, D., {Larsson}, J., {Fransson}, C., {et~al.} 2018, ArXiv e-prints.
\newblock \doarXiv{1805.04526}

\bibitem[{{Anders} \& {Grevesse}(1989)}]{anders89}
{Anders}, E., \& {Grevesse}, N. 1989, \gca, 53, 197,
  \dodoi{10.1016/0016-7037(89)90286-X}

\bibitem[{{Arnaud}(1996)}]{arnaud96}
{Arnaud}, K.~A. 1996, in Astronomical Society of the Pacific Conference Series,
  Vol. 101, Astronomical Data Analysis Software and Systems V, ed. G.~H.
  {Jacoby} \& J.~{Barnes}, 17

\bibitem[{{Arnett} {et~al.}(1989){Arnett}, {Bahcall}, {Kirshner}, \&
  {Woosley}}]{arnett89}
{Arnett}, W.~D., {Bahcall}, J.~N., {Kirshner}, R.~P., \& {Woosley}, S.~E. 1989,
  \araa, 27, 629, \dodoi{10.1146/annurev.aa.27.090189.003213}

\bibitem[{{Ashworth}(1980)}]{ashworth80}
{Ashworth}, Jr., W.~B. 1980, Journal for the History of Astronomy, 11, 1

\bibitem[{{Astropy Collaboration} {et~al.}(2013){Astropy Collaboration},
  {Robitaille}, {Tollerud}, {Greenfield}, {Droettboom}, {Bray}, {Aldcroft},
  {Davis}, {Ginsburg}, {Price-Whelan}, {Kerzendorf}, {Conley}, {Crighton},
  {Barbary}, {Muna}, {Ferguson}, {Grollier}, {Parikh}, {Nair}, {Unther},
  {Deil}, {Woillez}, {Conseil}, {Kramer}, {Turner}, {Singer}, {Fox}, {Weaver},
  {Zabalza}, {Edwards}, {Azalee Bostroem}, {Burke}, {Casey}, {Crawford},
  {Dencheva}, {Ely}, {Jenness}, {Labrie}, {Lim}, {Pierfederici}, {Pontzen},
  {Ptak}, {Refsdal}, {Servillat}, \& {Streicher}}]{astropy13}
{Astropy Collaboration}, {Robitaille}, T.~P., {Tollerud}, E.~J., {et~al.} 2013,
  \aap, 558, A33, \dodoi{10.1051/0004-6361/201322068}

\bibitem[{{Barlow} {et~al.}(2010){Barlow}, {Krause}, {Swinyard}, {Sibthorpe},
  {Besel}, {Wesson}, {Ivison}, {Dunne}, {Gear}, {Gomez}, {Hargrave}, {Henning},
  {Leeks}, {Lim}, {Olofsson}, \& {Polehampton}}]{barlow10}
{Barlow}, M.~J., {Krause}, O., {Swinyard}, B.~M., {et~al.} 2010, \aap, 518,
  L138, \dodoi{10.1051/0004-6361/201014585}

\bibitem[{{Barlow} {et~al.}(2013){Barlow}, {Swinyard}, {Owen}, {Cernicharo},
  {Gomez}, {Ivison}, {Krause}, {Lim}, {Matsuura}, {Miller}, {Olofsson}, \&
  {Polehampton}}]{barlow13}
{Barlow}, M.~J., {Swinyard}, B.~M., {Owen}, P.~J., {et~al.} 2013, Science, 342,
  1343, \dodoi{10.1126/science.1243582}

\bibitem[{{Bentley} {et~al.}(2018){Bentley}, {Wootten}, {Loh}, {Baldwin},
  {Ferland}, {Fabian}, {Combes}, {Salome}, \& {Castro-Carrizo}}]{bentley18}
{Bentley}, R., {Wootten}, A., {Loh}, E., {et~al.} 2018, in American
  Astronomical Society Meeting Abstracts, Vol. 231, American Astronomical
  Society Meeting Abstracts \#231, \#241.01

\bibitem[{{Bethe} \& {Salpeter}(1957)}]{bethe57}
{Bethe}, H.~A., \& {Salpeter}, E.~E. 1957, {Quantum Mechanics of One- and
  Two-Electron Atoms} ({New York: Academic Press, 1957})

\bibitem[{{Bionta} {et~al.}(1987){Bionta}, {Blewitt}, {Bratton}, {Casper}, \&
  {Ciocio}}]{bionta87}
{Bionta}, R.~M., {Blewitt}, G., {Bratton}, C.~B., {Casper}, D., \& {Ciocio}, A.
  1987, Physical Review Letters, 58, 1494, \dodoi{10.1103/PhysRevLett.58.1494}

\bibitem[{{Blinnikov} {et~al.}(2000){Blinnikov}, {Lundqvist}, {Bartunov},
  {Nomoto}, \& {Iwamoto}}]{blinnikov00}
{Blinnikov}, S., {Lundqvist}, P., {Bartunov}, O., {Nomoto}, K., \& {Iwamoto},
  K. 2000, \apj, 532, 1132, \dodoi{10.1086/308588}

\bibitem[{{Blondin} \& {Lundqvist}(1993)}]{blondin93}
{Blondin}, J.~M., \& {Lundqvist}, P. 1993, \apj, 405, 337,
  \dodoi{10.1086/172366}

\bibitem[{{Borkowski} {et~al.}(2010){Borkowski}, {Reynolds}, {Green}, {Hwang},
  {Petre}, {Krishnamurthy}, \& {Willett}}]{borkowski10}
{Borkowski}, K.~J., {Reynolds}, S.~P., {Green}, D.~A., {et~al.} 2010, \apjl,
  724, L161, \dodoi{10.1088/2041-8205/724/2/L161}

\bibitem[{{Bratton} {et~al.}(1988){Bratton}, {Casper}, {Ciocio}, {Claus},
  {Crouch}, {Dye}, {Errede}, {Gajewski}, {Goldhaber}, {Haines}, {Jones},
  {Kielczewska}, {Kropp}, {Learned}, {Losecco}, {Matthews}, {Miller}, {Mudan},
  {Price}, {Reines}, {Schultz}, {Seidel}, {Sinclair}, {Sobel}, {Stone},
  {Sulak}, {Svoboda}, {Thornton}, \& {van der Velde}}]{bratton88}
{Bratton}, C.~B., {Casper}, D., {Ciocio}, A., {et~al.} 1988, \prd, 37, 3361,
  \dodoi{10.1103/PhysRevD.37.3361}

\bibitem[{{Bruenn}(1993)}]{bruenn93}
{Bruenn}, S.~W. 1993, in Nuclear Physics in the Universe, ed. M.~W. {Guidry} \&
  M.~R. {Strayer}, 31--50

\bibitem[{{B{\"u}hler} \& {Blandford}(2014)}]{buhler14}
{B{\"u}hler}, R., \& {Blandford}, R. 2014, Reports on Progress in Physics, 77,
  066901, \dodoi{10.1088/0034-4885/77/6/066901}

\bibitem[{{Chakrabarty} {et~al.}(2001){Chakrabarty}, {Pivovaroff}, {Hernquist},
  {Heyl}, \& {Narayan}}]{chakrabarty01}
{Chakrabarty}, D., {Pivovaroff}, M.~J., {Hernquist}, L.~E., {Heyl}, J.~S., \&
  {Narayan}, R. 2001, \apj, 548, 800, \dodoi{10.1086/318994}

\bibitem[{{Chatterjee} \& {Cordes}(2002)}]{chatterjee02}
{Chatterjee}, S., \& {Cordes}, J.~M. 2002, \apj, 575, 407,
  \dodoi{10.1086/341139}

\bibitem[{{Chatterjee} \& {Cordes}(2004)}]{chatterjee04}
---. 2004, \apjl, 600, L51, \dodoi{10.1086/381498}

\bibitem[{{Cordes} {et~al.}(1993){Cordes}, {Romani}, \& {Lundgren}}]{cordes93}
{Cordes}, J.~M., {Romani}, R.~W., \& {Lundgren}, S.~C. 1993, \nat, 362, 133,
  \dodoi{10.1038/362133a0}

\bibitem[{{Davis} {et~al.}(2012){Davis}, {Bautz}, {Dewey}, {Heilmann}, {Houck},
  {Huenemoerder}, {Marshall}, {Nowak}, {Schattenburg}, {Schulz}, \&
  {Smith}}]{davis12}
{Davis}, J.~E., {Bautz}, M.~W., {Dewey}, D., {et~al.} 2012, in \procspie, Vol.
  8443, Space Telescopes and Instrumentation 2012: Ultraviolet to Gamma Ray,
  84431A

\bibitem[{{Dessart} \& {Hillier}(2010)}]{dessart10}
{Dessart}, L., \& {Hillier}, D.~J. 2010, \mnras, 405, 2141,
  \dodoi{10.1111/j.1365-2966.2010.16611.x}

\bibitem[{{Dickey} \& {Lockman}(1990)}]{dickey90}
{Dickey}, J.~M., \& {Lockman}, F.~J. 1990, \araa, 28, 215,
  \dodoi{10.1146/annurev.aa.28.090190.001243}

\bibitem[{{Dotani} {et~al.}(1987){Dotani}, {Hayashida}, {Inoue}, {Itoh},
  {Koyama}, {Makino}, {Mitsuda}, {Murakami}, {Oda}, {Ogawara}, {Takano},
  {Tanaka}, {Yoshida}, {Makishima}, {Ohashi}, {Kawai}, {Matsuoka}, {Hoshi},
  {Hayakawa}, {Kii}, {Kunieda}, {Nagase}, {Tawara}, {Hatsukade}, {Kitamoto},
  {Miyamoto}, {Tsunemi}, {Yamashita}, {Nakagawa}, {Yamauchi}, {Turner},
  {Pounds}, {Thomas}, {Stewart}, {Cruise}, {Patchett}, \& {Reading}}]{dotani87}
{Dotani}, T., {Hayashida}, K., {Inoue}, H., {et~al.} 1987, \nat, 330, 230,
  \dodoi{10.1038/330230a0}

\bibitem[{{Draine}(2003)}]{draine03}
{Draine}, B.~T. 2003, \apj, 598, 1026, \dodoi{10.1086/379123}

\bibitem[{{Dwek} \& {Arendt}(2015)}]{dwek15}
{Dwek}, E., \& {Arendt}, R.~G. 2015, \apj, 810, 75,
  \dodoi{10.1088/0004-637X/810/1/75}

\bibitem[{{Ebisuzaki} \& {Shibazaki}(1988)}]{ebisuzaki88}
{Ebisuzaki}, T., \& {Shibazaki}, N. 1988, \apjl, 327, L5,
  \dodoi{10.1086/185128}

\bibitem[{{Elshamouty} {et~al.}(2013){Elshamouty}, {Heinke}, {Sivakoff}, {Ho},
  {Shternin}, {Yakovlev}, {Patnaude}, \& {David}}]{elshamouty13}
{Elshamouty}, K.~G., {Heinke}, C.~O., {Sivakoff}, G.~R., {et~al.} 2013, \apj,
  777, 22, \dodoi{10.1088/0004-637X/777/1/22}

\bibitem[{{Esposito} {et~al.}(2018){Esposito}, {Rea}, {Lazzati}, {Matsuura},
  {Perna}, \& {Pons}}]{esposito18}
{Esposito}, P., {Rea}, N., {Lazzati}, D., {et~al.} 2018, \apj, 857, 58,
  \dodoi{10.3847/1538-4357/aab6b6}

\bibitem[{{Faucher-Gigu{\`e}re} \& {Kaspi}(2006)}]{faucher-giguere06}
{Faucher-Gigu{\`e}re}, C.-A., \& {Kaspi}, V.~M. 2006, \apj, 643, 332,
  \dodoi{10.1086/501516}

\bibitem[{{Fesen} {et~al.}(2006){Fesen}, {Hammell}, {Morse}, {Chevalier},
  {Borkowski}, {Dopita}, {Gerardy}, {Lawrence}, {Raymond}, \& {van den
  Bergh}}]{fesen06b}
{Fesen}, R.~A., {Hammell}, M.~C., {Morse}, J., {et~al.} 2006, \apj, 645, 283,
  \dodoi{10.1086/504254}

\bibitem[{{Flamsteed}(1725)}]{flamsteed25}
{Flamsteed}, J. 1725, {Historia Coelestis Britannicae, tribus Voluminibus
  contenta (1675-1689), (1689-1720), vol. 1, 2, 3} (London: H.~Meere; in folio;
  DCC.f.9, DCC.f.10, DCC.f.11)

\bibitem[{{Fransson} \& {Chevalier}(1987)}]{fransson87}
{Fransson}, C., \& {Chevalier}, R.~A. 1987, \apjl, 322, L15,
  \dodoi{10.1086/185028}

\bibitem[{{Fransson} {et~al.}(2013){Fransson}, {Larsson}, {Spyromilio},
  {Chevalier}, {Gr{\"o}ningsson}, {Jerkstrand}, {Leibundgut}, {McCray},
  {Challis}, {Kirshner}, {Kjaer}, {Lundqvist}, \& {Sollerman}}]{fransson13}
{Fransson}, C., {Larsson}, J., {Spyromilio}, J., {et~al.} 2013, \apj, 768, 88,
  \dodoi{10.1088/0004-637X/768/1/88}

\bibitem[{{Fruscione} {et~al.}(2006){Fruscione}, {McDowell}, {Allen},
  {Brickhouse}, {Burke}, {Davis}, {Durham}, {Elvis}, {Galle}, {Harris},
  {Huenemoerder}, {Houck}, {Ishibashi}, {Karovska}, {Nicastro}, {Noble},
  {Nowak}, {Primini}, {Siemiginowska}, {Smith}, \& {Wise}}]{fruscione06}
{Fruscione}, A., {McDowell}, J.~C., {Allen}, G.~E., {et~al.} 2006, in
  \procspie, Vol. 6270, Society of Photo-Optical Instrumentation Engineers
  (SPIE) Conference Series, 62701V

\bibitem[{{Fryxell} {et~al.}(1991){Fryxell}, {Arnett}, \&
  {Mueller}}]{fryxell91}
{Fryxell}, B., {Arnett}, D., \& {Mueller}, E. 1991, \apj, 367, 619,
  \dodoi{10.1086/169657}

\bibitem[{{Garc{\'{\i}}a} {et~al.}(2005){Garc{\'{\i}}a}, {Mendoza}, {Bautista},
  {Gorczyca}, {Kallman}, \& {Palmeri}}]{garcia05}
{Garc{\'{\i}}a}, J., {Mendoza}, C., {Bautista}, M.~A., {et~al.} 2005, \apjs,
  158, 68, \dodoi{10.1086/428712}

\bibitem[{{Garc{\'{\i}}a} {et~al.}(2009){Garc{\'{\i}}a}, {Kallman},
  {Witthoeft}, {Behar}, {Mendoza}, {Palmeri}, {Quinet}, {Bautista}, \&
  {Klapisch}}]{garcia09}
{Garc{\'{\i}}a}, J., {Kallman}, T.~R., {Witthoeft}, M., {et~al.} 2009, \apjs,
  185, 477, \dodoi{10.1088/0067-0049/185/2/477}

\bibitem[{{Gatuzz} {et~al.}(2015){Gatuzz}, {Garc{\'{\i}}a}, {Kallman},
  {Mendoza}, \& {Gorczyca}}]{gatuzz15}
{Gatuzz}, E., {Garc{\'{\i}}a}, J., {Kallman}, T.~R., {Mendoza}, C., \&
  {Gorczyca}, T.~W. 2015, \apj, 800, 29, \dodoi{10.1088/0004-637X/800/1/29}

\bibitem[{{Gomez} {et~al.}(2012){Gomez}, {Krause}, {Barlow}, {Swinyard},
  {Owen}, {Clark}, {Matsuura}, {Gomez}, {Rho}, {Besel}, {Bouwman}, {Gear},
  {Henning}, {Ivison}, {Polehampton}, \& {Sibthorpe}}]{gomez12}
{Gomez}, H.~L., {Krause}, O., {Barlow}, M.~J., {et~al.} 2012, \apj, 760, 96,
  \dodoi{10.1088/0004-637X/760/1/96}

\bibitem[{{Gorczyca}(2000)}]{gorczyca00}
{Gorczyca}, T.~W. 2000, \pra, 61, 024702, \dodoi{10.1103/PhysRevA.61.024702}

\bibitem[{{Gorczyca} {et~al.}(2013){Gorczyca}, {Bautista}, {Hasoglu},
  {Garc{\'{\i}}a}, {Gatuzz}, {Kaastra}, {Kallman}, {Manson}, {Mendoza},
  {Raassen}, {de Vries}, \& {Zatsarinny}}]{gorczyca13}
{Gorczyca}, T.~W., {Bautista}, M.~A., {Hasoglu}, M.~F., {et~al.} 2013, \apj,
  779, 78, \dodoi{10.1088/0004-637X/779/1/78}

\bibitem[{{Grebenev} \& {Syunyaev}(1988)}]{grebenev88}
{Grebenev}, S.~A., \& {Syunyaev}, R.~A. 1988, Soviet Astronomy Letters, 14, 288

\bibitem[{{Grevesse} \& {Sauval}(1998)}]{grevesse98}
{Grevesse}, N., \& {Sauval}, A.~J. 1998, \ssr, 85, 161,
  \dodoi{10.1023/A:1005161325181}

\bibitem[{{Haso{\u g}lu} {et~al.}(2014){Haso{\u g}lu}, {Abdel-Naby}, {Gatuzz},
  {Garc{\'{\i}}a}, {Kallman}, {Mendoza}, \& {Gorczyca}}]{hasoglu14}
{Haso{\u g}lu}, M.~F., {Abdel-Naby}, S.~A., {Gatuzz}, E., {et~al.} 2014, \apjs,
  214, 8, \dodoi{10.1088/0067-0049/214/1/8}

\bibitem[{{Heinke} \& {Ho}(2010)}]{heinke10}
{Heinke}, C.~O., \& {Ho}, W.~C.~G. 2010, \apjl, 719, L167,
  \dodoi{10.1088/2041-8205/719/2/L167}

\bibitem[{{HI4PI Collaboration} {et~al.}(2016){HI4PI Collaboration}, {Ben
  Bekhti}, {Fl{\"o}er}, {Keller}, {Kerp}, {Lenz}, {Winkel}, {Bailin},
  {Calabretta}, {Dedes}, {Ford}, {Gibson}, {Haud}, {Janowiecki}, {Kalberla},
  {Lockman}, {McClure-Griffiths}, {Murphy}, {Nakanishi}, {Pisano}, \&
  {Staveley-Smith}}]{hi4pi16}
{HI4PI Collaboration}, {Ben Bekhti}, N., {Fl{\"o}er}, L., {et~al.} 2016, \aap,
  594, A116, \dodoi{10.1051/0004-6361/201629178}

\bibitem[{{Hirata} {et~al.}(1987){Hirata}, {Kajita}, {Koshiba}, {Nakahata}, \&
  {Oyama}}]{hirata87}
{Hirata}, K., {Kajita}, T., {Koshiba}, M., {Nakahata}, M., \& {Oyama}, Y. 1987,
  Physical Review Letters, 58, 1490, \dodoi{10.1103/PhysRevLett.58.1490}

\bibitem[{{Hirata} {et~al.}(1988){Hirata}, {Kajita}, {Koshiba}, {Nakahata},
  {Oyama}, {Sato}, {Suzuki}, {Takita}, {Totsuka}, {Kifune}, {Suda},
  {Takahashi}, {Tanimori}, {Miyano}, {Yamada}, {Beier}, {Feldscher}, {Frati},
  {Kim}, {Mann}, {Newcomer}, {van Berg}, {Zhang}, \& {Cortez}}]{hirata88}
{Hirata}, K.~S., {Kajita}, T., {Koshiba}, M., {et~al.} 1988, \prd, 38, 448,
  \dodoi{10.1103/PhysRevD.38.448}

\bibitem[{{Ho} \& {Heinke}(2009)}]{ho09}
{Ho}, W.~C.~G., \& {Heinke}, C.~O. 2009, \nat, 462, 71,
  \dodoi{10.1038/nature08525}

\bibitem[{{Hobbs} {et~al.}(2005){Hobbs}, {Lorimer}, {Lyne}, \&
  {Kramer}}]{hobbs05}
{Hobbs}, G., {Lorimer}, D.~R., {Lyne}, A.~G., \& {Kramer}, M. 2005, \mnras,
  360, 974, \dodoi{10.1111/j.1365-2966.2005.09087.x}

\bibitem[{{Hughes}(1980)}]{hughes80}
{Hughes}, D.~W. 1980, \nat, 285, 132, \dodoi{10.1038/285132a0}

\bibitem[{{Hunter}(2007)}]{hunter07}
{Hunter}, J.~D. 2007, Computing in Science and Engineering, 9, 90,
  \dodoi{10.1109/MCSE.2007.55}

\bibitem[{{Indebetouw} {et~al.}(2014){Indebetouw}, {Matsuura}, {Dwek},
  {Zanardo}, {Barlow}, {Baes}, {Bouchet}, {Burrows}, {Chevalier}, {Clayton},
  {Fransson}, {Gaensler}, {Kirshner}, {Laki{\'c}evi{\'c}}, {Long}, {Lundqvist},
  {Mart{\'{\i}}-Vidal}, {Marcaide}, {McCray}, {Meixner}, {Ng}, {Park},
  {Sonneborn}, {Staveley-Smith}, {Vlahakis}, \& {van Loon}}]{indebetouw14}
{Indebetouw}, R., {Matsuura}, M., {Dwek}, E., {et~al.} 2014, \apjl, 782, L2,
  \dodoi{10.1088/2041-8205/782/1/L2}

\bibitem[{{Jerkstrand} {et~al.}(2011){Jerkstrand}, {Fransson}, \&
  {Kozma}}]{jerkstrand11}
{Jerkstrand}, A., {Fransson}, C., \& {Kozma}, C. 2011, \aap, 530, A45,
  \dodoi{10.1051/0004-6361/201015937}

\bibitem[{Jones {et~al.}(2001--)Jones, Oliphant, Peterson, {et~al.}}]{jones01}
Jones, E., Oliphant, T., Peterson, P., {et~al.} 2001--

\bibitem[{{Juett} {et~al.}(2006){Juett}, {Schulz}, {Chakrabarty}, \&
  {Gorczyca}}]{juett06}
{Juett}, A.~M., {Schulz}, N.~S., {Chakrabarty}, D., \& {Gorczyca}, T.~W. 2006,
  \apj, 648, 1066, \dodoi{10.1086/506189}

\bibitem[{{Kageyama} \& {Sato}(2004)}]{kageyama04}
{Kageyama}, A., \& {Sato}, T. 2004, Geochemistry, Geophysics, Geosystems, 5,
  Q09005, \dodoi{10.1029/2004GC000734}

\bibitem[{{Kalberla} {et~al.}(2005){Kalberla}, {Burton}, {Hartmann}, {Arnal},
  {Bajaja}, {Morras}, \& {P{\"o}ppel}}]{kalberla05}
{Kalberla}, P.~M.~W., {Burton}, W.~B., {Hartmann}, D., {et~al.} 2005, \aap,
  440, 775, \dodoi{10.1051/0004-6361:20041864}

\bibitem[{{Kamenetzky} {et~al.}(2013){Kamenetzky}, {McCray}, {Indebetouw},
  {Barlow}, {Matsuura}, {Baes}, {Blommaert}, {Bolatto}, {Decin}, {Dunne},
  {Fransson}, {Glenn}, {Gomez}, {Groenewegen}, {Hopwood}, {Kirshner},
  {Lakicevic}, {Marcaide}, {Marti-Vidal}, {Meixner}, {Royer}, {Soderberg},
  {Sonneborn}, {Staveley-Smith}, {Swinyard}, {Van de Steene}, {van Hoof}, {van
  Loon}, {Yates}, \& {Zanardo}}]{kamenetzky13}
{Kamenetzky}, J., {McCray}, R., {Indebetouw}, R., {et~al.} 2013, \apjl, 773,
  L34, \dodoi{10.1088/2041-8205/773/2/L34}

\bibitem[{{Kamper}(1980)}]{kamper80}
{Kamper}, K.~W. 1980, The Observatory, 100, 3

\bibitem[{{Keohane} {et~al.}(1996){Keohane}, {Rudnick}, \&
  {Anderson}}]{keohane96}
{Keohane}, J.~W., {Rudnick}, L., \& {Anderson}, M.~C. 1996, \apj, 466, 309,
  \dodoi{10.1086/177511}

\bibitem[{{Kifonidis} {et~al.}(2003){Kifonidis}, {Plewa}, {Janka}, \&
  {M{\"u}ller}}]{kifonidis03}
{Kifonidis}, K., {Plewa}, T., {Janka}, H.-T., \& {M{\"u}ller}, E. 2003, \aap,
  408, 621, \dodoi{10.1051/0004-6361:20030863}

\bibitem[{{Kirshner} {et~al.}(1987){Kirshner}, {Sonneborn}, {Crenshaw}, \&
  {Nassiopoulos}}]{kirshner87}
{Kirshner}, R.~P., {Sonneborn}, G., {Crenshaw}, D.~M., \& {Nassiopoulos}, G.~E.
  1987, \apj, 320, 602, \dodoi{10.1086/165579}

\bibitem[{{Kortright} \& {Kim}(2000)}]{kortright00}
{Kortright}, J.~B., \& {Kim}, S.-K. 2000, \prb, 62, 12216,
  \dodoi{10.1103/PhysRevB.62.12216}

\bibitem[{{Krause} {et~al.}(2008){Krause}, {Birkmann}, {Usuda}, {Hattori},
  {Goto}, {Rieke}, \& {Misselt}}]{krause08}
{Krause}, O., {Birkmann}, S.~M., {Usuda}, T., {et~al.} 2008, Science, 320,
  1195, \dodoi{10.1126/science.1155788}

\bibitem[{{Kuiper} \& {Hermsen}(2015)}]{kuiper15}
{Kuiper}, L., \& {Hermsen}, W. 2015, \mnras, 449, 3827,
  \dodoi{10.1093/mnras/stv426}

\bibitem[{{Leising}(2001)}]{leising01}
{Leising}, M.~D. 2001, \apj, 563, 185, \dodoi{10.1086/323776}

\bibitem[{{Leising}(2006)}]{leising06}
---. 2006, \apj, 651, 1019, \dodoi{10.1086/507602}

\bibitem[{{Limongi} {et~al.}(2000){Limongi}, {Straniero}, \&
  {Chieffi}}]{limongi00}
{Limongi}, M., {Straniero}, O., \& {Chieffi}, A. 2000, \apjs, 129, 625,
  \dodoi{10.1086/313424}

\bibitem[{{Long} {et~al.}(2012){Long}, {Blair}, {Godfrey}, {Kuntz},
  {Plucinsky}, {Soria}, {Stockdale}, {Whitmore}, \& {Winkler}}]{long12}
{Long}, K.~S., {Blair}, W.~P., {Godfrey}, L.~E.~H., {et~al.} 2012, \apj, 756,
  18, \dodoi{10.1088/0004-637X/756/1/18}

\bibitem[{{Matsuura} {et~al.}(2011){Matsuura}, {Dwek}, {Meixner}, {Otsuka},
  {Babler}, {Barlow}, {Roman-Duval}, {Engelbracht}, {Sandstrom},
  {Laki{\'c}evi{\'c}}, {van Loon}, {Sonneborn}, {Clayton}, {Long}, {Lundqvist},
  {Nozawa}, {Gordon}, {Hony}, {Panuzzo}, {Okumura}, {Misselt}, {Montiel}, \&
  {Sauvage}}]{matsuura11}
{Matsuura}, M., {Dwek}, E., {Meixner}, M., {et~al.} 2011, Science, 333, 1258,
  \dodoi{10.1126/science.1205983}

\bibitem[{{Matsuura} {et~al.}(2015){Matsuura}, {Dwek}, {Barlow}, {Babler},
  {Baes}, {Meixner}, {Cernicharo}, {Clayton}, {Dunne}, {Fransson}, {Fritz},
  {Gear}, {Gomez}, {Groenewegen}, {Indebetouw}, {Ivison}, {Jerkstrand},
  {Lebouteiller}, {Lim}, {Lundqvist}, {Pearson}, {Roman-Duval}, {Royer},
  {Staveley-Smith}, {Swinyard}, {van Hoof}, {van Loon}, {Verstappen}, {Wesson},
  {Zanardo}, {Blommaert}, {Decin}, {Reach}, {Sonneborn}, {Van de Steene}, \&
  {Yates}}]{matsuura15}
{Matsuura}, M., {Dwek}, E., {Barlow}, M.~J., {et~al.} 2015, \apj, 800, 50,
  \dodoi{10.1088/0004-637X/800/1/50}

\bibitem[{{Matsuura} {et~al.}(2017){Matsuura}, {Indebetouw}, {Woosley},
  {Bujarrabal}, {Abell{\'a}n}, {McCray}, {Kamenetzky}, {Fransson}, {Barlow},
  {Gomez}, {Cigan}, {De Looze}, {Spyromilio}, {Staveley-Smith}, {Zanardo},
  {Roche}, {Larsson}, {Viti}, {van Loon}, {Wheeler}, {Baes}, {Chevalier},
  {Lundqvist}, {Marcaide}, {Dwek}, {Meixner}, {Ng}, {Sonneborn}, \&
  {Yates}}]{matsuura17}
{Matsuura}, M., {Indebetouw}, R., {Woosley}, S., {et~al.} 2017, \mnras, 469,
  3347, \dodoi{10.1093/mnras/stx830}

\bibitem[{{Matz} {et~al.}(1988){Matz}, {Share}, {Leising}, {Chupp}, \&
  {Vestrand}}]{matz88}
{Matz}, S.~M., {Share}, G.~H., {Leising}, M.~D., {Chupp}, E.~L., \& {Vestrand},
  W.~T. 1988, \nat, 331, 416, \dodoi{10.1038/331416a0}

\bibitem[{{McCray}(1993)}]{mccray93}
{McCray}, R. 1993, \araa, 31, 175, \dodoi{10.1146/annurev.aa.31.090193.001135}

\bibitem[{{McCray} \& {Fransson}(2016)}]{mccray16}
{McCray}, R., \& {Fransson}, C. 2016, \araa, 54, 19,
  \dodoi{10.1146/annurev-astro-082615-105405}

\bibitem[{{Menon} \& {Heger}(2017)}]{menon17}
{Menon}, A., \& {Heger}, A. 2017, \mnras, 469, 4649,
  \dodoi{10.1093/mnras/stx818}

\bibitem[{{Morris} \& {Podsiadlowski}(2007)}]{morris07}
{Morris}, T., \& {Podsiadlowski}, P. 2007, Science, 315, 1103,
  \dodoi{10.1126/science.1136351}

\bibitem[{{Morrison} \& {McCammon}(1983)}]{morrison83}
{Morrison}, R., \& {McCammon}, D. 1983, \apj, 270, 119, \dodoi{10.1086/161102}

\bibitem[{{Morse} {et~al.}(2004){Morse}, {Fesen}, {Chevalier}, {Borkowski},
  {Gerardy}, {Lawrence}, \& {van den Bergh}}]{morse04}
{Morse}, J.~A., {Fesen}, R.~A., {Chevalier}, R.~A., {et~al.} 2004, \apj, 614,
  727, \dodoi{10.1086/423709}

\bibitem[{{M{\"u}ller} {et~al.}(2017){M{\"u}ller}, {Melson}, {Heger}, \&
  {Janka}}]{muller17}
{M{\"u}ller}, B., {Melson}, T., {Heger}, A., \& {Janka}, H.-T. 2017, \mnras,
  472, 491, \dodoi{10.1093/mnras/stx1962}

\bibitem[{{M{\"u}ller} {et~al.}(1991){M{\"u}ller}, {Fryxell}, \&
  {Arnett}}]{muller91}
{M{\"u}ller}, E., {Fryxell}, B., \& {Arnett}, D. 1991, \aap, 251, 505

\bibitem[{{Nomoto} \& {Hashimoto}(1988)}]{nomoto88}
{Nomoto}, K., \& {Hashimoto}, M. 1988, \physrep, 163, 13,
  \dodoi{10.1016/0370-1573(88)90032-4}

\bibitem[{{Orlando} {et~al.}(2015){Orlando}, {Miceli}, {Pumo}, \&
  {Bocchino}}]{orlando15}
{Orlando}, S., {Miceli}, M., {Pumo}, M.~L., \& {Bocchino}, F. 2015, \apj, 810,
  168, \dodoi{10.1088/0004-637X/810/2/168}

\bibitem[{{Owen} \& {Barlow}(2015)}]{owen15}
{Owen}, P.~J., \& {Barlow}, M.~J. 2015, \apj, 801, 141,
  \dodoi{10.1088/0004-637X/801/2/141}

\bibitem[{{{\"O}zel} \& {Freire}(2016)}]{ozel16}
{{\"O}zel}, F., \& {Freire}, P. 2016, \araa, 54, 401,
  \dodoi{10.1146/annurev-astro-081915-023322}

\bibitem[{{Panagia}(1999)}]{panagia99}
{Panagia}, N. 1999, in IAU Symposium, Vol. 190, New Views of the Magellanic
  Clouds, ed. Y.-H. {Chu}, N.~{Suntzeff}, J.~{Hesser}, \& D.~{Bohlender}, 549

\bibitem[{{Panagia} {et~al.}(1991){Panagia}, {Gilmozzi}, {Macchetto}, {Adorf},
  \& {Kirshner}}]{panagia91}
{Panagia}, N., {Gilmozzi}, R., {Macchetto}, F., {Adorf}, H.-M., \& {Kirshner},
  R.~P. 1991, \apjl, 380, L23, \dodoi{10.1086/186164}

\bibitem[{{Pavlov} {et~al.}(2000){Pavlov}, {Zavlin}, {Aschenbach},
  {Tr{\"u}mper}, \& {Sanwal}}]{pavlov00}
{Pavlov}, G.~G., {Zavlin}, V.~E., {Aschenbach}, B., {Tr{\"u}mper}, J., \&
  {Sanwal}, D. 2000, \apjl, 531, L53, \dodoi{10.1086/312521}

\bibitem[{{Posselt} {et~al.}(2013){Posselt}, {Pavlov}, {Suleimanov}, \&
  {Kargaltsev}}]{posselt13}
{Posselt}, B., {Pavlov}, G.~G., {Suleimanov}, V., \& {Kargaltsev}, O. 2013,
  \apj, 779, 186, \dodoi{10.1088/0004-637X/779/2/186}

\bibitem[{{Rampp} \& {Janka}(2002)}]{rampp02}
{Rampp}, M., \& {Janka}, H.-T. 2002, \aap, 396, 361,
  \dodoi{10.1051/0004-6361:20021398}

\bibitem[{{Reed} {et~al.}(1995){Reed}, {Hester}, {Fabian}, \&
  {Winkler}}]{reed95}
{Reed}, J.~E., {Hester}, J.~J., {Fabian}, A.~C., \& {Winkler}, P.~F. 1995,
  \apj, 440, 706, \dodoi{10.1086/175308}

\bibitem[{{Rest} {et~al.}(2011){Rest}, {Foley}, {Sinnott}, {Welch}, {Badenes},
  {Filippenko}, {Bergmann}, {Bhatti}, {Blondin}, {Challis}, {Damke}, {Finley},
  {Huber}, {Kasen}, {Kirshner}, {Matheson}, {Mazzali}, {Minniti}, {Nakajima},
  {Narayan}, {Olsen}, {Sauer}, {Smith}, \& {Suntzeff}}]{rest11}
{Rest}, A., {Foley}, R.~J., {Sinnott}, B., {et~al.} 2011, \apj, 732, 3,
  \dodoi{10.1088/0004-637X/732/1/3}

\bibitem[{{Rho} {et~al.}(2012){Rho}, {Onaka}, {Cami}, \& {Reach}}]{rho12}
{Rho}, J., {Onaka}, T., {Cami}, J., \& {Reach}, W.~T. 2012, \apjl, 747, L6,
  \dodoi{10.1088/2041-8205/747/1/L6}

\bibitem[{{Rho} {et~al.}(2008){Rho}, {Kozasa}, {Reach}, {Smith}, {Rudnick},
  {DeLaney}, {Ennis}, {Gomez}, \& {Tappe}}]{rho08}
{Rho}, J., {Kozasa}, T., {Reach}, W.~T., {et~al.} 2008, \apj, 673, 271,
  \dodoi{10.1086/523835}

\bibitem[{{Rybicki} \& {Lightman}(1979)}]{rybicki79}
{Rybicki}, G.~B., \& {Lightman}, A.~P. 1979, {Radiative processes in
  astrophysics} (New York, Wiley-Interscience)

\bibitem[{{Saio} {et~al.}(1988){Saio}, {Nomoto}, \& {Kato}}]{saio88}
{Saio}, H., {Nomoto}, K., \& {Kato}, M. 1988, \nat, 334, 508,
  \dodoi{10.1038/334508a0}

\bibitem[{{Salas} {et~al.}(2018){Salas}, {Oonk}, {van Weeren}, {Wolfire},
  {Emig}, {Toribio}, {R{\"o}ttgering}, \& {Tielens}}]{salas18}
{Salas}, P., {Oonk}, J.~B.~R., {van Weeren}, R.~J., {et~al.} 2018, \mnras, 475,
  2496, \dodoi{10.1093/mnras/stx3340}

\bibitem[{{Scheck} {et~al.}(2006){Scheck}, {Kifonidis}, {Janka}, \&
  {M{\"u}ller}}]{scheck06}
{Scheck}, L., {Kifonidis}, K., {Janka}, H.-T., \& {M{\"u}ller}, E. 2006, \aap,
  457, 963, \dodoi{10.1051/0004-6361:20064855}

\bibitem[{{Serafimovich} {et~al.}(2004){Serafimovich}, {Shibanov}, {Lundqvist},
  \& {Sollerman}}]{serafimovich04}
{Serafimovich}, N.~I., {Shibanov}, Y.~A., {Lundqvist}, P., \& {Sollerman}, J.
  2004, \aap, 425, 1041, \dodoi{10.1051/0004-6361:20040499}

\bibitem[{{Shigeyama} \& {Nomoto}(1990)}]{shigeyama90}
{Shigeyama}, T., \& {Nomoto}, K. 1990, \apj, 360, 242, \dodoi{10.1086/169114}

\bibitem[{{Shtykovskiy} {et~al.}(2005){Shtykovskiy}, {Lutovinov}, {Gilfanov},
  \& {Sunyaev}}]{shtykovskiy05}
{Shtykovskiy}, P.~E., {Lutovinov}, A.~A., {Gilfanov}, M.~R., \& {Sunyaev},
  R.~A. 2005, Astronomy Letters, 31, 258, \dodoi{10.1134/1.1896069}

\bibitem[{{Stage} {et~al.}(2004){Stage}, {Joss}, {Madej}, \&
  {R{\'o}{\.z}a{\'n}ska}}]{stage04}
{Stage}, M.~D., {Joss}, P.~C., {Madej}, J., \& {R{\'o}{\.z}a{\'n}ska}, A. 2004,
  Advances in Space Research, 33, 605, \dodoi{10.1016/j.asr.2003.08.026}

\bibitem[{{Suleimanov} {et~al.}(2014){Suleimanov}, {Klochkov}, {Pavlov}, \&
  {Werner}}]{suleimanov14}
{Suleimanov}, V.~F., {Klochkov}, D., {Pavlov}, G.~G., \& {Werner}, K. 2014,
  \apjs, 210, 13, \dodoi{10.1088/0067-0049/210/1/13}

\bibitem[{{Sunyaev} {et~al.}(1987){Sunyaev}, {Kaniovsky}, {Efremov},
  {Gilfanov}, {Churazov}, {Grebenev}, {Kuznetsov}, {Melioranskiy},
  {Yamburenko}, {Yunin}, {Stepanov}, {Chulkov}, {Pappe}, {Boyarskiy},
  {Gavrilova}, {Loznikov}, {Prudkoglyad}, {Rodin}, {Reppin}, {Pietsch},
  {Engelhauser}, {Truemper}, {Voges}, {Kendziorra}, {Bezler}, {Staubert},
  {Brinkman}, {Heise}, {Mels}, {Jager}, {Skinner}, {Al-Emam}, {Patterson},
  {Willmore}, {Gilfanov}, \& {Churazov}}]{sunyaev87}
{Sunyaev}, R., {Kaniovsky}, A., {Efremov}, V., {et~al.} 1987, \nat, 330, 227,
  \dodoi{10.1038/330227a0}

\bibitem[{{Tananbaum}(1999)}]{tananbaum99}
{Tananbaum}, H. 1999, \iaucirc, 7246

\bibitem[{{Theiling} \& {Leising}(2006)}]{theiling06}
{Theiling}, M.~F., \& {Leising}, M.~D. 2006, \nar, 50, 544,
  \dodoi{10.1016/j.newar.2006.06.054}

\bibitem[{{Truelove} \& {McKee}(1999)}]{truelove99}
{Truelove}, J.~K., \& {McKee}, C.~F. 1999, \apjs, 120, 299,
  \dodoi{10.1086/313176}

\bibitem[{{Urushibata} {et~al.}(2018){Urushibata}, {Takahashi}, {Umeda}, \&
  {Yoshida}}]{urushibata18}
{Urushibata}, T., {Takahashi}, K., {Umeda}, H., \& {Yoshida}, T. 2018, \mnras,
  473, L101, \dodoi{10.1093/mnrasl/slx166}

\bibitem[{{Utrobin} {et~al.}(2015){Utrobin}, {Wongwathanarat}, {Janka}, \&
  {M{\"u}ller}}]{utrobin15}
{Utrobin}, V.~P., {Wongwathanarat}, A., {Janka}, H.-T., \& {M{\"u}ller}, E.
  2015, \aap, 581, A40, \dodoi{10.1051/0004-6361/201425513}

\bibitem[{{Utrobin} {et~al.}(2018)}]{utrobin18}
{Utrobin}, V.~P., {et~al.} 2018, in preparation

\bibitem[{van~der Walt {et~al.}(2011)van~der Walt, Colbert, \&
  Varoquaux}]{van_der_walt11}
van~der Walt, S., Colbert, S.~C., \& Varoquaux, G. 2011, Computing in Science
  Engineering, 13, 22, \dodoi{10.1109/MCSE.2011.37}

\bibitem[{{Verner} {et~al.}(1996){Verner}, {Ferland}, {Korista}, \&
  {Yakovlev}}]{verner96}
{Verner}, D.~A., {Ferland}, G.~J., {Korista}, K.~T., \& {Yakovlev}, D.~G. 1996,
  \apj, 465, 487, \dodoi{10.1086/177435}

\bibitem[{{Verner} \& {Yakovlev}(1995)}]{verner95}
{Verner}, D.~A., \& {Yakovlev}, D.~G. 1995, \aaps, 109, 125

\bibitem[{{Walborn} {et~al.}(1987){Walborn}, {Lasker}, {Laidler}, \&
  {Chu}}]{walborn87}
{Walborn}, N.~R., {Lasker}, B.~M., {Laidler}, V.~G., \& {Chu}, Y.-H. 1987,
  \apjl, 321, L41, \dodoi{10.1086/185002}

\bibitem[{{West} {et~al.}(1987){West}, {Lauberts}, {Schuster}, \&
  {Jorgensen}}]{west87}
{West}, R.~M., {Lauberts}, A., {Schuster}, H.-E., \& {Jorgensen}, H.~E. 1987,
  \aap, 177, L1

\bibitem[{{White} \& {Malin}(1987)}]{white87}
{White}, G.~L., \& {Malin}, D.~F. 1987, \nat, 327, 36, \dodoi{10.1038/327036a0}

\bibitem[{{Wilms} {et~al.}(2000){Wilms}, {Allen}, \& {McCray}}]{wilms00}
{Wilms}, J., {Allen}, A., \& {McCray}, R. 2000, \apj, 542, 914,
  \dodoi{10.1086/317016}

\bibitem[{{Winkel} {et~al.}(2016){Winkel}, {Kerp}, {Fl{\"o}er}, {Kalberla},
  {Ben Bekhti}, {Keller}, \& {Lenz}}]{winkel16}
{Winkel}, B., {Kerp}, J., {Fl{\"o}er}, L., {et~al.} 2016, \aap, 585, A41,
  \dodoi{10.1051/0004-6361/201527007}

\bibitem[{{Witthoeft} {et~al.}(2009){Witthoeft}, {Bautista}, {Mendoza},
  {Kallman}, {Palmeri}, \& {Quinet}}]{witthoeft09}
{Witthoeft}, M.~C., {Bautista}, M.~A., {Mendoza}, C., {et~al.} 2009, \apjs,
  182, 127, \dodoi{10.1088/0067-0049/182/1/127}

\bibitem[{{Witthoeft} {et~al.}(2011){Witthoeft}, {Garc{\'{\i}}a}, {Kallman},
  {Bautista}, {Mendoza}, {Palmeri}, \& {Quinet}}]{witthoeft11}
{Witthoeft}, M.~C., {Garc{\'{\i}}a}, J., {Kallman}, T.~R., {et~al.} 2011,
  \apjs, 192, 7, \dodoi{10.1088/0067-0049/192/1/7}

\bibitem[{{Wongwathanarat} {et~al.}(2010{\natexlab{a}}){Wongwathanarat},
  {Hammer}, \& {M{\"u}ller}}]{wongwathanarat10b}
{Wongwathanarat}, A., {Hammer}, N.~J., \& {M{\"u}ller}, E. 2010{\natexlab{a}},
  \aap, 514, A48, \dodoi{10.1051/0004-6361/200913435}

\bibitem[{{Wongwathanarat} {et~al.}(2010{\natexlab{b}}){Wongwathanarat},
  {Janka}, \& {M{\"u}ller}}]{wongwathanarat10}
{Wongwathanarat}, A., {Janka}, H.-T., \& {M{\"u}ller}, E. 2010{\natexlab{b}},
  \apjl, 725, L106, \dodoi{10.1088/2041-8205/725/1/L106}

\bibitem[{{Wongwathanarat} {et~al.}(2013){Wongwathanarat}, {Janka}, \&
  {M{\"u}ller}}]{wongwathanarat13}
---. 2013, \aap, 552, A126, \dodoi{10.1051/0004-6361/201220636}

\bibitem[{{Wongwathanarat} {et~al.}(2017){Wongwathanarat}, {Janka},
  {M{\"u}ller}, {Pllumbi}, \& {Wanajo}}]{wongwathanarat17}
{Wongwathanarat}, A., {Janka}, H.-T., {M{\"u}ller}, E., {Pllumbi}, E., \&
  {Wanajo}, S. 2017, \apj, 842, 13, \dodoi{10.3847/1538-4357/aa72de}

\bibitem[{{Wongwathanarat} {et~al.}(2015){Wongwathanarat}, {M{\"u}ller}, \&
  {Janka}}]{wongwathanarat15}
{Wongwathanarat}, A., {M{\"u}ller}, E., \& {Janka}, H.-T. 2015, \aap, 577, A48,
  \dodoi{10.1051/0004-6361/201425025}

\bibitem[{{Woosley} {et~al.}(1988){Woosley}, {Pinto}, \& {Ensman}}]{woosley88}
{Woosley}, S.~E., {Pinto}, P.~A., \& {Ensman}, L. 1988, \apj, 324, 466,
  \dodoi{10.1086/165908}

\bibitem[{{Woosley} \& {Weaver}(1995)}]{woosley95}
{Woosley}, S.~E., \& {Weaver}, T.~A. 1995, \apjs, 101, 181,
  \dodoi{10.1086/192237}

\bibitem[{{Xu} {et~al.}(1988){Xu}, {Sutherland}, {McCray}, \& {Ross}}]{xu88}
{Xu}, Y., {Sutherland}, P., {McCray}, R., \& {Ross}, R.~R. 1988, \apj, 327,
  197, \dodoi{10.1086/166181}

\end{thebibliography}
\end{document}